\documentclass[review]{elsarticle}

\usepackage{lineno,hyperref}
\usepackage{amsmath}
\usepackage{graphicx}
\modulolinenumbers[5]
\usepackage[ruled,vlined]{algorithm2e}
\numberwithin{equation}{section}
\usepackage{algpseudocode}
\usepackage{makecell}
\usepackage{adjustbox}
\usepackage[figuresright]{rotating}
\usepackage{makecell}
\usepackage{setspace}
\usepackage{color}
\usepackage{cases}

\journal{Journal of \LaTeX\ Templates}

\newdefinition{rmk}{Definition}
\DeclareMathOperator*{\argmax}{arg\,max}

\begin{document}

\begin{frontmatter}

\title{Game Theory and Machine Learning in UAVs-Assisted Wireless Communication Networks: A Survey\\}
\author[rvt]{M. Zhou \corref{cor1} \fnref{fn1}}
\ead{mzhou91@gatech.edu}
\author[focal]{Y. Guan}
\author[els]{M. Hayajneh}
\author[rvt]{K. Niu}
\author[rvt]{C. Abdallah}

\cortext[cor1]{Corresponding author }

\address[rvt]{Department of Electrical and Computer Engineering, Georgia Institute of Technology, Atlanta, GA, 30332, USA}
\address[focal]{Department
of Aerospace Engineering, Georgia Institute of Technology, Atlanta, GA, 30332, USA}
\address[els]{Department
of Computer and Network Engineering, United Arab Emirates University, P. O. Box 15551, Al Ain}
\fntext[fn1]{Mailing address: 333638 Georgia Tech Station, Atlanta, GA, 30332, USA}

\begin{abstract}
In recent years, Unmanned Aerial Vehicles (UAVs) have been used in fields such as architecture, business delivery, military and civilian theaters, and many others.
With increased applications comes the increased demand for advanced algorithms for resource allocation and energy management. 
As is well known, game theory and machine learning are two powerful tools already
widely used in the wireless communication field and there are numerous surveys of game theory and machine learning usage in wireless communication.
Existing surveys however focus either on game theory or machine learning and due to this fact, the current article surveys both game-theoretic and machine learning algorithms for use by UAVs in Wireless Communication Networks (U-WCNs). 
We also discuss how to combine game theory and machine learning for solving problems in U-WCNs and identify several future research directions.
\end{abstract}


\end{frontmatter}
\linenumbers

\section{Introduction}
Unmanned Aerial Vehicles (UAVs) are increasingly being deployed in wireless communication networks, largely due to their low cost and unrestricted mobility~\cite{tutorialsofUAV}.
Notable usage examples include the Google Loon project~\cite{googleloon} and the Facebook Aquila project~\cite{facebookAquila}. 
In these examples, UAVs serve as mobile Base Stations (BS) directly providing wireless communication for users, or as relays between devices and fixed base stations. 
UAVs-assisted networks have also found applications in fields that require reliable communication or assured identity~\cite{huaweiwhitepaper} such as precision agriculture, search and rescue, and parcel delivery as discussed next. 

With the recent boom in the number of mobile and Internet-of-Things (IoT) devices, swarms of UAVs may be needed to assist in establishing communication networks~\cite{huaweiwhitepaper,huaweivideo}.
For example, in precision agriculture, multiple UAVs are deployed to assist  in irrigation management, crops health monitoring, and cattle herding.  
These are labor-intensive tasks due to the dense distribution of crops and the continuous mobility of animals.
The advantages of swarm UAVs in such cases include time savings and cost reduction~\cite{swarm_UAVs_review}.
In search and rescue tasks, a swarm of UAVs are able to work cooperatively in extremely harsh disaster environments. 
UAVs are able to quickly and efficiently search an area, identifying victims and their status, then communicating such information to ground assets~\cite{SARescue_UAV}.
Autonomous driving is also benefiting from advances in U-WCNs.
As an example, Vehicle-to-Everything (V2X) communication systems will allow vehicles to connect to everything using UAVs, which can act either as a medium of data transmission between vehicles and base stations or as security enhancers~\cite{V2X}. 
Finally, one of the most immediate applications of swarm UAVs is for delivery service~\cite{AmazonDelivery}.
In this scenario, the UAVs help deliver packages to customers' backyards, and rendezvous with delivery trucks.
All these applications rely not only on the safe flight control of each UAV but also on their ability to communicate wirelessly and reliably.

Traditionally, deploying UAVs in wireless communications systems faced challenges such as complicated channels models~\cite{surveychannel,zhanghanbook}, dynamic cell association~\cite{zhanghanbook}, energy constraints~\cite{zhanghanbook} and legislative regulations~\cite{Khamvilai2021}.
With the continuing increase of the number of deployed UAVs, new challenges associated with multi-agent decision making also arise. 
Such challenges include multi-agent trajectory planning~\cite{MFG_NN}, multi-agent resource allocation~\cite{jointaccessselect,DDPG-RA, MFQ} and user association~\cite{MILP_cluster}. 

Game theory provides tools to solve multi-agent decision problems and to analyze the interactions among various agents in a communication network.  
Game theoretic concepts such as Nash or correlated equilibrium are well suited for U-WCNs~\cite{owen:Game-Theory}.
With the increasing number of UAVs required to accomplish complex tasks however, traditional game theoretic algorithms may become intractable. 
One possible approach to tackling this challenge is to leverage machine learning techniques such as function approximation~\cite{neuralDynamicProgramming, Mnih2015_DQN}, policy gradient~\cite{DDPG}, and multi-agent actor critic~\cite{RL_sutton, lowe2017multi}.
While abundant literature exists for game theory~\cite{survey-UAV-2,GT_survey} and machine learning~\cite{DRL_wireless,ML-UAV-survey} approaches to U-WCNs problems, few if any exist with a unified treatment of the two areas. 

This survey attempts to fill this void by first reviewing the existing literature, then providing linkages between game theoretic and machine learning techniques for UAVs-assisted wireless communication systems.

\subsection{Prior surveys}
There are many surveys of UAVs-assisted wireless communication networks~\cite{tutorialsofUAV,civil,router, fanet,fotouhi2018survey,surveychannel,LAP_survey}.
The authors of~\cite{civil} for example, reported on the characteristics and requirements of UAV networks for multiple civilian applications.
These include search and rescue, area coverage (e.g., monitoring and surveillance), network coverage (e.g., relays/base stations/data mule), delivery, and construction. 
In particular, the Quality-of-Service (QoS) requirements, network-relevant mission parameters, data requirements, connectivity, adaptability, safety, and privacy were discussed.
Reference \cite{fotouhi2018survey} covered a variety of cellular-specific issues such as Third Generation Partnership Project (3GPP) development, vendor prototypes, regulations, and cyber-security issues that affect the cellular UAVs and potential business model.
The authors also proposed multiple future research directions such as UAV simulators, advanced UAV mobility control based on image processing and deep learning, new antenna designs to achieve higher data rate, physical reliability, and mobile edge computing.
Reference~\cite{router} discussed some important issues in UAVs communication networks, such as the characteristics of UAV networks and the protocols in various layers to assist in greening the network.
The authors compared the advantages and disadvantages of various network structures (e.g., star vs mesh), different routing protocols (e.g., static, proactive, on-demand, or reactive, hybrid), and existing seamless handovers.
Reference~\cite{surveychannel} presented a comprehensive and unified review of UAVs' air-to-ground channel models.
Reference \cite{fanet} focused on applications of Flying Ad-hoc Networks (FANETs) based on UAVs, such as traffic monitoring, agricultural management, military defense, and relay networks. 
Furthermore, the authors considered the communication challenges in FANETs systems. 
These challenges include high mobility, frequent topology changes, minimal delay, and high reliability requirements. 
Reference \cite{LAP_survey} offered an overall view of High Altitude Platform (HAP)-based and Low Altitude Platform (LAP)-based communication networks, as well as Airborne Communication Networks (ACN).
Reference \cite{tutorialsofUAV} presented a description of the potential applications and benefits of UAVs in wireless communication networks.  
It briefly described using game theory, Machine Learning (ML), and optimization theory to solve certain challenges in U-WCNs,  such as Three-Dimensional (3D) deployment and energy optimization. 

Earlier research has focused on connecting game theory and wireless communications. 
As an example, Reference~\cite{survey-UAV-2} presented a number of game-theoretic solutions for energy consumption optimization, network coverage enhancement, and connectivity improvement in wireless communication systems using UAVs. 
In particular, the authors proposed Mean-Field Games (MFG) to solve problems in \textit{massive} UAVs networks. 
Reference \cite{GT_survey} utilized game-theoretic tools to model and analyze UAVs-assisted networks, where various problems within the physical layer, the data link layer, the network layer, the transport layer, and the application layer, were modeled and studied using various game formulations such as potential games, Bayesian games, and mean field games.

Likewise, there are many surveys describing the use of machine learning methods in conventional wireless communication systems and UAVs-assisted wireless communication networks. 
Reference \cite{DRL_wireless} provided a comprehensive review of Deep Reinforcement Learning (DRL) in communication and networking. 
The authors reviewed recent DRL methods addressing issues such as dynamic network access, data rate control, wireless caching, data offloading, network security, and connectivity preservation. 
That review, however, only briefly touched upon the recent applications of DRL in UAVs. 
In~\cite{ML-UAV-survey}, the authors provided an overview of ML techniques in U-WCNs, such as propagation channel modeling, resource management, security, and positioning. 
Other open issues for ML applications in UAVs-based networks are also identified in both the networking and security areas. 
Reference \cite{ML_UAV_intro} listed several applications of  ML techniques (e.g., supervised learning and reinforcement learning) in UAV-based Radio Access Networks (RAN).
These applications include radio resource allocation, design of collectors and relays, choice of the type and number of UAVs, positioning of UAVs acting as BSs, and the design of a mobile cloud. 

We summarize the contributions of these surveys in Table \ref{table:relevantSurvey}.
Note that to the best of our knowledge, none of the previous surveys for U-WCNs have dealt with the intersection of machine learning and game theory.
With the increasing interest in wireless communication applications requiring a large number of UAVs, ours seems to be the first survey that presents a unified view of the two fields.

\subsection{Game theory and machine learning in UAVs-assisted wireless communication networks}
\begin{figure}[htbp] 
\centerline{\includegraphics[width=\textwidth]{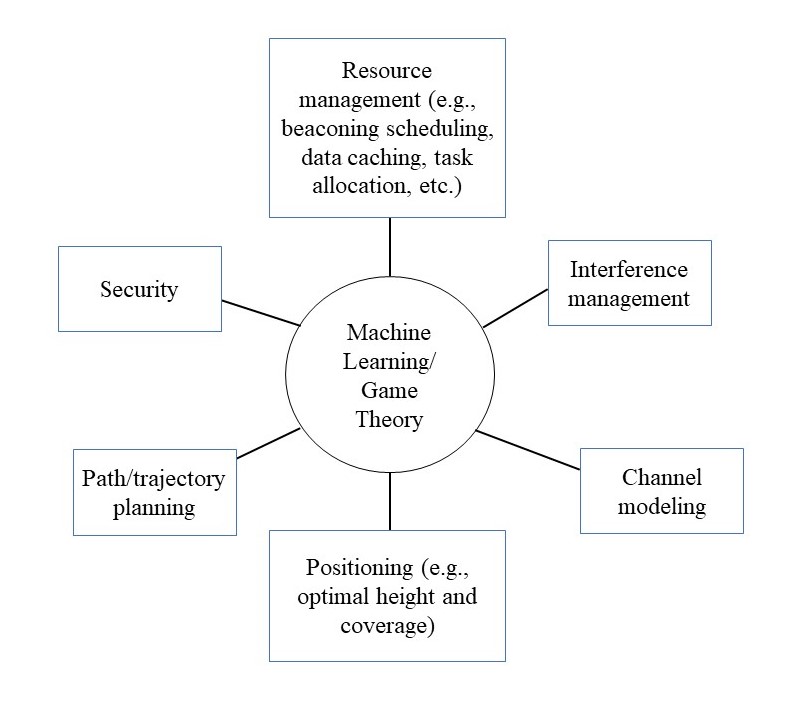}}
\caption{Scenarios of machine learning and game theory in UAVs-assisted wireless communication networks.}
\label{fig:ML_GT_application}
\end{figure}

Game theory and machine learning are two pillars that support applications in UAVs-assisted wireless communication networks.
Notable examples include resource management~\cite{zhanghanbook,jointaccessselect,coalition,Charging,Koulali2016AGS,POCA,UAV_offloading,QL,double_Qlearning,DDPG-RA}, positioning~\cite{zhang2021multiagent,QL-emergency,non-coop-coverage}, trajectory planning~\cite{zhanghanbook,zhang2021multiagent,MFG-movementcontrol}, interference management~\cite{SG_Antijamming_Bayesian,Attack-PT-Q}, channel modeling~\cite{ML-UAV-survey,channelmodeling} and security~\cite{jamming,BayesianGame,SG_Antijamming_Bayesian,Attack-PT-Q}. 
Fig.~\ref{fig:ML_GT_application} presents various applications of machine learning and game theory in U-WCNs.
We give next a brief introduction to each of the application, and present a more detailed discussion in later sections.

\textbf{Positioning~\cite{zhang2021multiagent,QL-emergency,non-coop-coverage}:} The height and elevation angle of a UAV impact its coverage performance and link reliability over a service area \cite{zhanghanbook}.
Furthermore, the optimal density of UAVs in an area is subject to safety and interference constraints.
Research related to UAV positioning focuses on maximizing the coverage of the system while minimizing the interference.
 
\textbf{Path/trajectory planning~\cite{zhanghanbook,zhang2021multiagent,MFG-movementcontrol}:}
Subject to energy limitation, the trajectories of UAVs in a network need to be optimized, with link quality, interference and collision avoidance taken into consideration.

\textbf{Security~\cite{jamming,BayesianGame,SG_Antijamming_Bayesian,Attack-PT-Q}:} 
Jamming and eavesdropping between UAVs and devices are two major security problems in U-WCNs.
Both induce huge economical and political losses to companies and users.

\textbf{Resource management~\cite{zhanghanbook,jointaccessselect,coalition,Charging,Koulali2016AGS,POCA,UAV_offloading,QL,double_Qlearning,DDPG-RA}:} Mobile devices and IoT devices have limited battery lifetime and constrained storage capability. 
As a result, in a UAV-cellular network, the UAVs need to support data caching and content relaying. 
Each UAV may be assigned different tasks (caching or relaying) and may also select different users to serve. 
The objective of resource management is to maximize the revenue of the operator(s), by optimizing the task assignments and user selection. 
Furthermore, if the UAVs belong to different operators, competition among the operators also need to be considered. 

\textbf{Interference management~\cite{SG_Antijamming_Bayesian,Attack-PT-Q}:} Interference exists in both traditional terrestrial networks and UAVs-assisted networks.
For the latter, the interference comes from three sources: other communicating UAVs, mobile users, and ground control stations.

\textbf{Channel modeling~\cite{ML-UAV-survey,channelmodeling}:} Working in a 3D dynamic environment, UAVs have to operate in a more complex channel model that accounts for the weather, obstacles, and the Doppler shift effect.

\subsection{Our contribution}
As described earlier, there are many surveys of the application of game theory and machine learning methods to vehicular networks \cite{survey-V2V}, smart grids \cite{survey-smartGrid} and wireless sensor networks \cite{WSN-survey}. To the best of our knowledge, such surveys focus on either the machine learning ~\cite{DRL_wireless, ML-UAV-survey,ML_UAV_intro} or the game theory tools~\cite{survey-UAV-2, GT_survey}.
The present survey attempts to provide the first unified survey that connects these two well-studied areas with their applications in U-WCNs.  
Rather than simply combining existing surveys, we examine the intrinsic connections between game theory, machine learning, and their applications to U-WCNs.

\begin{sidewaystable}[htbp]
\caption{Relevant surveys and magazines in UAVs-assisted wireless communication networks (N = No, Y = Yes, B = Brief introduction).}
\vspace{+10pt}
\resizebox{\textwidth}{!}{
\begin{tabular}
{c|c|c|c|c }
\hline
\textbf{References} & \textbf{Topics} &\makecell[c]{\textbf{Game} \\[-2pt] \textbf{Theory}} &
\makecell[c]{\textbf{Machine}\\[-2pt] \textbf{Learning}} & \makecell[c]{\textbf{Potential}\\[-2pt] \textbf{Challenges}} \\ \hline
  \cite{civil}  & \makecell[c]{Characteristics and requirements\\[-2pt] of UAV networks}  &   N&  N & Y\\ \hline
\cite{router}  & \makecell[c]{Characteristics, routing, \\[-2pt] handover scheme in UAV networks}  & N & N & Y\\ \hline
  \cite{surveychannel}  & Air-to-ground channel model   &  N & N & Y\\ \hline
\cite{fanet}  & Applications and challenges of FANETs & N  & N  &Y \\ \hline
\cite{fotouhi2018survey}  &  \makecell[c]{Standardization, regulations,\\[-2pt] security, future direction} &  N& N& Y  \\ \hline
 \cite{LAP_survey} &  LAP, HAP, ACN  & N & N & Y\\ \hline 
 \cite{tutorialsofUAV} &\makecell[c]{ Opportunities, challenges,\\[-2pt] open problems, and mathematical tools} & B & B & Y\\ \hline
 \cite{survey-UAV-2} & \makecell[c]{Game theoretic solutions for energy,\\[-2pt] coverage optimization, task allocation, etc.}& Y  & N & Y\\ \hline
\cite{GT_survey}  & \makecell[c]{Game theoretic tools for modeling\\[-2pt] and analyzing UAV-assisted networks} & N & N & Y\\ \hline
\cite{DRL_wireless}  & DRL in communications and networking & N & Y &Y \\ \hline
\cite{ML-UAV-survey} & ML applications in UAV-based networks & N & Y &Y \\ \hline
\cite{ML_UAV_intro} & ML in UAV-based RAN &  N& Y & Y\\ \hline \makecell[c]{Our survey} &

\makecell[c]{
Game theory and machine learning \\ techniques in UAVs-assisted wireless\\[-2pt] communication, challenges and solutions }
  & Y  & Y  & Y  \\ \hline
\end{tabular}
}
\label{table:relevantSurvey}
\end{sidewaystable}
\subsection{Organization}
The remainder of this article is organized as follows. In Section~\ref{sec:application}, we discuss the potential applications and challenges of UAVs for wireless communication. 
In Section~\ref{sec:GT}, we present some game theoretic techniques used to analyze wireless communication systems with UAVs. 
In Section~\ref{sec:ML}, we introduce machine learning algorithms for UAVs-assisted wireless communication systems.
In Section \ref{sec:intersection}, we discuss the intersection of game theory and machine learning for U-WCNs, and present open problems then list several promising research directions.
Section \ref{sec:conclusion} concludes this survey.
\section{Wireless communication with UAVs: motivating applications and challenges}\label{sec:application}
Depending on their flying altitude, UAVs are categorized into high-altitude platforms ($>17$ km) and low-altitude platforms. 
The low-altitude platforms have the advantages of higher flexibility, lower cost, lower latency, and easier maintenance, making them more suitable for Fifth-Generation wireless (5G) and IoT services. 
High-altitude platforms on the other hand, provide a more sustainable wireless network coverage for rural environments.
This article focuses on low-altitude platforms and more specifically on unmanned aerial drones.

\begin{figure}[htbp]
\centerline{\includegraphics[width=\textwidth]{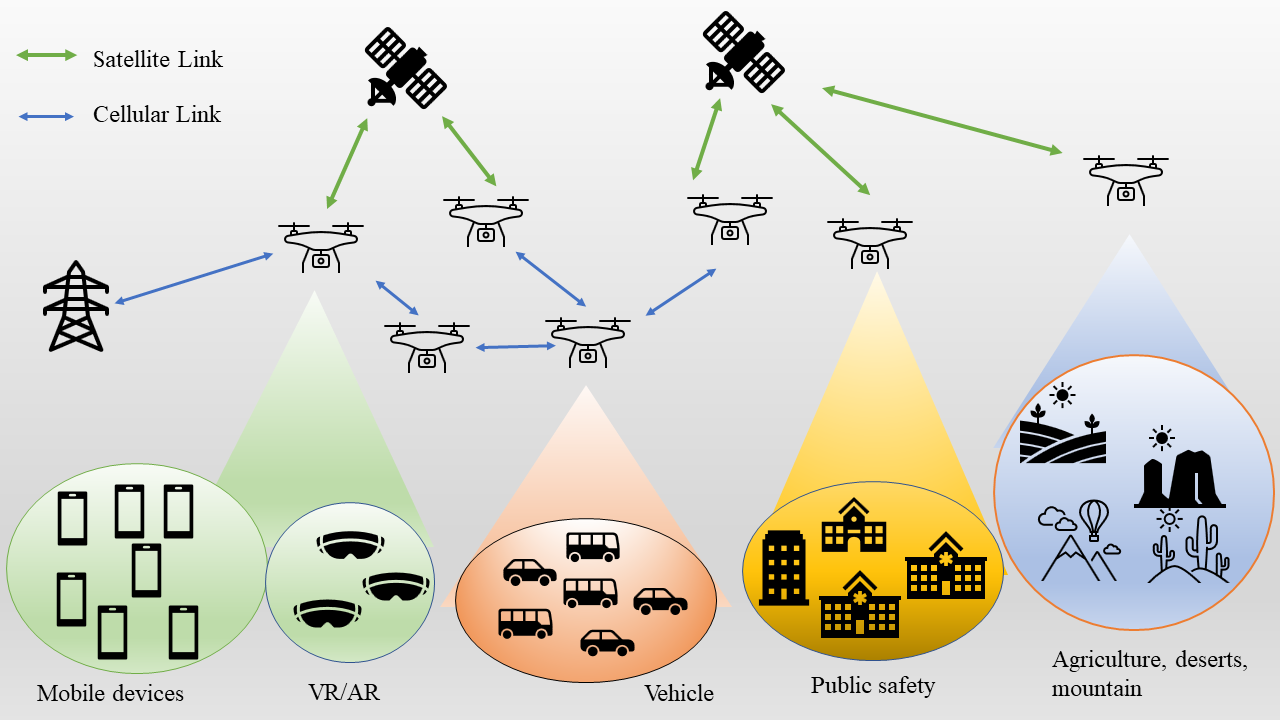}}
\caption{Applications of UAVs-assisted networks.}
 \label{fig:UAV-applications}
\end{figure}

UAVs play different roles in various wireless communication settings. Fig.~\ref{fig:UAV-applications} shows some of those roles in future 5G and IoT networks. 
On one hand, UAVs may be used as aerial base stations in the 5G and beyond eras.
Such UAVs improve the reliability of wireless links in Device-to-Device (D2D) and Vehicle-to-Vehicle (V2V) communications. 
On the other hand, aerial platforms are suitable for maintaining fast and ubiquitous connectivity whenever ground wireless networks fail after natural disasters~\cite{naturaldisaster}. 
UAVs can also serve as relays for communication among base stations and user devices. 
In addition, UAVs may be flying users as part of a cellular network for delivery applications and in Virtual Reality (VR) / Augmented Reality (AR) situations, where UAVs capture desired information about a specific area and transmit it to remote users in real time \cite{tutorialsofUAV}. 
In summary, UAVs can boost the performance of existing ground wireless networks in terms of coverage, capacity, delay and overall quality of service. 

Despite the ubiquity of their potential applications, many challenges remain for the wide deployment of UAVs. 
The first is the complexity of the UAVs-user channel model. 
The Air-to-Ground (A2G) channels are susceptible to blockage and affected by weather, altitude, elevation angle, type of UAVs, and propagation environments. For the A2G channel modeling problem as an example, there exists no specific modeling method for channel measurements in urban areas and rural areas under various weather conditions. 
In a dynamic UAV-to-UAV communication network, channel modelling is also complicated by the time-varying nature of the channel and the Doppler effect.
The second challenge is the deployment and trajectory optimization problem.
When integrating UAVs into communication systems, one would like to minimize the transmission latency of users, minimize the energy consumption while simultaneously maximizing the spectral efficiency and coverage performance. 
As a result, it is necessary to optimize the locations and the trajectories of UAVs, as well as the bandwidth/power allocation among them. Thus a framework that can dynamically manage these various resources while keeping the interference to ground users at acceptable levels is needed. Of course, the interference from UAVs to ground users should also be addressed.
In addition, UAVs that act as users within cellular networks require a dynamic handover mechanism design and new scheduling schemes. 
Finally, as the use cases of UAVs increase (e.g., online video streaming, medical delivery), various security challenges may arise. For example, an attacker may disrupt the UAV's data transmission or send malicious data causing irregular movement and collisions, ultimately resulting in significant losses \cite{security}.

Trying to address the 3D location and trajectory design problem, Reference \cite{tutorialsofUAV} proposed using convex optimization and optimal transport theory. 
Reference \cite{UAV_op} presented a framework to jointly optimize the 3D placement and mobility of UAVs, device-UAV association, and uplink power control.
This framework breaks the complicated optimization problem into two sub-problems and solves the sub-problems in an iterative manner. 
Many similar problems are transformed into simpler but still challenging mixed integer programming problems, which either can not be solved by conventional optimization methods due to their non-convexity, or may still require high computational resources.

Game Theory (GT) methods were introduced to assist in the modeling and solution of the optimization problem. 
Game theory provides a solid foundation for distributed decision making in UAVs-assisted wireless networks.
In a game-theoretic framework, UAVs, BSs, and User Equipments (UEs) are regarded as players in a game, while the energy, spectrum, 3D positions and flight times are considered as the strategy spaces.
This allows us to frame the optimization problem using existing machinery developed for stochastic differential games, coalitional games, mean-field games, contract theory, and others. 

With the development of high performance computing hardware and the availability of large data sets, Machine Learning (ML) techniques have recently been applied to many fields due to their ability of ``learning" from interacting with the environment. 
For UAVs-assisted wireless communication systems, ML algorithms enable UAVs to promptly adjust their positions, trajectories, flight directions, and motion control to serve the ground users.
Moreover, ML algorithms may also be used to build a 3D channel model for UAVs~\cite{tutorialsofUAV}.
Further synergies with optimization theory and game theory enlarge the range of problems that machine learning can address in UAVs-assisted wireless communication systems. 
For example, Reference \cite{MILP_cluster} combined Mixed Integer Linear Programming (MILP) with clustering methods to maximize the weighted sum rate of UAV-served users and the total number of D2D-connected users. 
This method reduces the time complexity of solving such problems while maintaining as good performance as the classical MILP methods.

In the following two sections, we present a detailed summary of game theory and machine learning techniques in the field of UAVs-assisted wireless communication networks and some state-of-the-art algorithms.
\section{Game theory in UAVs-assisted wireless communication} \label{sec:GT}
Game theory studies the strategic interactions among rational players. 
More specifically, it deals with problems where multiple rational players interact with each other strategically in order to maximize their own benefit.
Unlike most traditional optimization methods, game theory often provides efficient and robust distributed algorithms and has thus found extensive applications in wireless networks for modeling, analyzing, and designing distributed schemes~\cite{GT_survey,GT-wireless}.

For UAVs-assisted wireless communication systems, one needs to resolve the load balancing, offloading, and distributed resource management problems among UAVs, BSs, and UEs. 
On the other hand, trade-offs between energy, spectrum, and 3D locations also require attention.
In this article, we focus on game theoretic concepts and methods to solve both problems in U-WCNs as described next.

In general, a game~\cite{GT1991} is composed of three elements: the set of players denoted by ${\mathcal{N}=\{1,2,...,i,...,n\}}$, the strategy space for each player $i$ denoted by ${S_i=\{s_1,s_2,...,s_m}\}$, and the payoff function $u_i$ also known as the reward that players receive at the end of the game contingent upon the actions of all other players in the game.
In a UAVs-assisted wireless communication network, the players may be UAVs, ground users, or base stations. 
The strategies may be the beaconing periods scheduling, task servicing, UAVs relocating, offloading, channel assigning, and intruders evading.  
The payoff may be chosen as the throughput, Signal-to-Interference-plus-Noise Ratio (SINR), delays, or the number of nodes covered based on real applications~\cite{GT_survey}.
A game is static if all players make decisions simultaneously without knowledge of other players' strategies. 
It is dynamic when the players make decisions sequentially or repeatedly. 
Based on whether the information structure is known or not, games may be divided into two categories: complete-information games and incomplete-information games.
In addition, a game is characterized as a perfect-information or imperfect-information game based on whether all players know the historical actions of each other when they take their actions.
Based on whether the players are cooperating to optimize a common goal or not, games can also be divided into cooperative games and non-cooperative games. 
The following list gives the definition of additional terms in the game theory literature:
\begin{itemize}
    \item Stochastic game~\cite{Stochasticgame}: The game moves to a new state governed by transition probabilities that depend on the previous state and actions taken. 
    The total payoff is defined as the discounted cumulative rewards of the payoffs during the course of the game.
    \item  Nash equilibrium:  When all players are operating at the Nash equilibrium, any unilateral deviation of an agent from this equilibrium point would not improve that agent's total payoff. 
    A formal definition of a Nash equilibrium is: \\
    \begin{rmk} (Nash equilibrium \cite{zhanghanbook}): We denote an action profile of the players as $a=\{ a_1, a_2, ..., a_M\}$. 
    An action profile $a^* = \{ a_1^*, a_2^*,..., a_M^* \}$ is a pure-strategy Nash Equilibrium (NE) if and only if no player could improve its utility $u_m$ by deviating unilaterally, i.e.,
\begin{equation}
    u_m (a^*_m, a^*_{-m}) \geq u_m(a_m, a_  {-m}^*) \quad  \text{for any action $a_m$}.
\end{equation}
\end{rmk}
\end{itemize}

\begin{figure}[htbp] 
\centerline{\includegraphics[width=\textwidth]{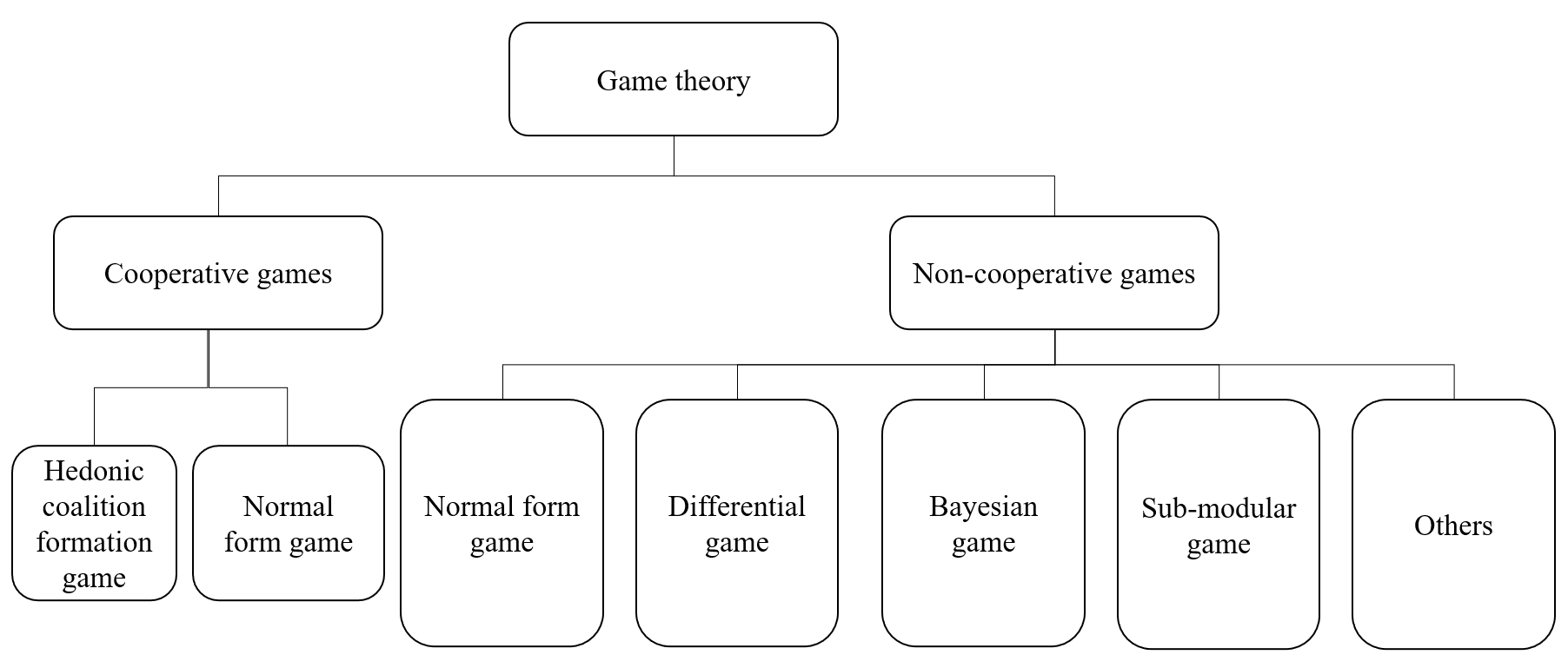}}
\caption{Classification of current game theoretic approaches used in UAVs-assisted wireless communications~\cite{GT_survey_simply}.}
\label{fig:GT_classification}
\end{figure}

In the following subsection, we introduce game theoretic concepts and their corresponding applications in U-WCNs.
A more detailed description of game theory may be found in \cite{GT_survey} and \cite{GT_survey_simply}.
Fig.~\ref{fig:GT_classification} presents a general classification of classical game-theoretic approaches used in U-WCNs.

\subsection{Cooperative games}
A cooperative game, also known as coalitional game, is a game where the players form coalitions and take joint action as a group. 
The players within each group cooperate with each other while the group members with the same objective compete with members of other groups.
In disaster scenarios, for example, UAVs have an incentive to cooperatively provide alternative network access to users in order to reduce their loss.
UAVs from the same operator would like to cooperatively take on tasks (e.g. routing, data collecting, etc.) in order to maximize the revenues of the operator. 
Thus, cooperative game theory may be used to model the problems of rate allocation, cooperative transmission, packet forwarding, and so on.
The drawback of the coalitional game however, is that it is NP-complete, i.e., with the increase of the size of the communication network, the time complexity increases.
Thus, numerous heuristic algorithms are usually used to find a near-optimal solution in large communication networks.

Hedonic coalitional formation game is a special class of coalitional formation game where the players are self-interested and only care about the identity of the players in their coalition, and each player has a preference rank over different coalitions.
In~\cite{coalition}, a number of UAVs are required to collect data from several arbitrarily-located tasks in a UAV-based flying ad-hoc network. 
A hedonic coalitional formation game is used to model the interactions between UAVs and tasks in order to form disjoint coalitions. 
Both the tasks and the UAVs are players who decide to join or leave a coalition based on their payoffs. 
The total utility of every coalition is evaluated using a coalitional value function defined as the ratio of some power of the throughput and delay.
Each formed coalition is modeled as a polling system comprised of a number of UAVs that move between different tasks to collect and transmit packets to a common receiver. 
Considering the computational complexity, the UAVs operate based on the nearest neighbor route.
The coalition keeps updating until a Nash stable network partition is reached.
The authors compared the performance of this algorithm with the algorithm that assigns the tasks equally among UAVs.
The simulation results show that the proposed algorithm outperforms equal allocation in terms of the average payoff by at least 30\% no matter how many tasks there are.

In game theory, a normal-form game is a game that is represented by a matrix, as opposed to the extensive form representation. 
In a normal-form game, pure strategies may not exist while a Nash equilibrium is guaranteed to exist in a mixed strategy which is the probability distribution over pure strategies for a player.
This formulation is useful in identifying strictly dominant strategies (a strictly dominant strategy is one that always provides greater utility to the player, independent of the other player's strategy) and Nash equilibrium strategies and has gained popularity in wireless communication applications. 
A downside of the normal-form game formulation however, is the potential loss of some information.
Such information includes the sequencing of agents' probable moves, their possible strategies at every decision-making point, the inadequate information each agent has about the other agents’ moves when they make a decision, and their payoffs for all possible game outcomes.

A mixed strategy normal-form game is used for the recharging schedule in ~\cite{Charging}, where the authors proposed a joint coverage, connectivity, and charging strategy mechanism for a mesh of UAVs.   
The UAVs aim to maximize the stationary coverage of a target area, while simultaneously guaranteeing the continuity of service with necessary recharging. 
In this formulation, the scheduling of recharging operations is considered as a set of consecutive and different static normal-form games at each time slot $t_i$.
The players are all the UAVs; the action set contains the following elements:~\\
\{access the replenishment station ($G_{\text{OK}}$), remain in state $s_{\text{fly}}$ ($G_{\text{NO}}$), release the replenishment station and change state to $s_{\text{fly}}$ ($R_{\text{OK}}$), remain in state recharging ($R_{\text{NO}}$)\}.~\\ 
The payoff is the energy defined as a function of the state-action pair.
Then the mixed strategy is achieved when the expected utility is indifferent of its possible choice, i.e., $u(R_{\text{OK}}) = u(R_{\text{NO}})$.
In the experiment, the authors compared the system life time and failed recharge attempt ratio in three cases, i.e., global knowledge (know all other UAVs' residue energy), local knowledge (only know the UAVs' energy at one-hop distance) and personal knowledge (only know its own residue energy) with the centralized coordination algorithm and probability approach.
The results show that the game theory-based solutions outperform the probability approach but slightly under-perform the centralized coordination solution.

\subsection{Non-cooperative games}
As opposed to cooperative games, non-cooperative game theory deals with the scenario when individual players compete with each other to maximize their own payoffs.   
This type of game therefore assumes that all players are self-interested.
There exist various kinds of non-cooperative games, such as differential games, Bayesian games, sub-modular games, and so on.
Non-cooperative game theory is more commonly used to model competing relationships in UAVs-assisted wireless communication for power control, resource allocation, positioning of UAVs and security.
For example, two UAVs belonging to different operators compete for business \cite{non-coop-coverage,Koulali2016AGS,POCA,UAV_offloading} and military UAVs try to monitor, jam or anti-jam enemy's communication systems~\cite{jamming, BayesianGame}. 

Reference \cite{non-coop-coverage} studied the positioning problem of UAVs in order to maximize the coverage of mobile devices (i.e., the number of mobile devices connected).
In this case, the mobile devices are randomly moving on the ground.
Three UAVs choose to either circle in their current cell or move to circle the center of an adjacent cell based on the number of mobile devices it supports.
The payoff matrix contains the values for the coverage of each UAV if the corresponding action is selected.
Then all players choose their strategies simultaneously by finding one Nash equilibrium from this payoff matrix.
The coverage of UAVs is shown to improve by 11.9\%.
This game theoretic scheme is also shown to be more energy-efficient compared to the three single-UAV coverage scenario.
However, only three UAVs are considered in that article and it is well-known that the normal-form game has a scalability problem when the number of players increases.
Furthermore, when comparing the power efficiency, only communication power is considered while the power needed for movement is not taken into consideration.

Reference~\cite{Koulali2016AGS} focused on the beaconing scheduling problem between two non-cooperative UAVs.  
The two UAVs belong to different operators and independently optimize their beaconing period to provide coverage for the mobile users on the ground.
This problem is formulated as a sub-modular game where UAVs are the players and they strategically choose their beaconing schedule.
The payoff of each UAV taking beaconing strategy profile $(\tau_i,\tau_j)$ is defined as a function of the  encounter rate and energy consumption, which is
\begin{align}
    u_i^i(\tau_i,\tau_j) = P_s^i(\tau_i,\tau_j)-\frac{(C_b\tau_i+C_s)l}{m},
\end{align}
where $m=l\times T$ is the available time window for UAVs to contact with mobile devices, $T$ is the beaconing period, $l$ is a constant, $P_s^i(\tau_i,\tau_j)$ is the successful encounter rate, $C_b$ and $C_s$ are respectively the energy cost per slot for sending beacons and energy cost for remaining switching the transceiver state.
Due to the special property of this payoff function (sub-modular function), a pure strategy Nash equilibrium exists with the assumption of perfect rationality and complete knowledge.
To overcome this limitation, the authors provided an adaptive distributed learning framework based on the ``Nash Seeking Algorithm (NSA)"~\cite{book-NSA} to find the Nash equilibrium.
The advantage of this distributed algorithm is that each UAV strategy is only based on its own observations and the exact formula of the payoff is not even needed.
To verify the efficacy of the proposed NSA algorithm, simulation results are provided to show that the algorithm converges to the same value but slightly slower than the Best Response Dynamics (BRD) algorithm.

In \cite{POCA}, the authors used a non-cooperative model to explore the radio channels assignment problems in a combined UAV and D2D-based networks.
These assignment problems are generally challenging due to the limited availability of orthogonal channels, interference, dynamic topology, and the high mobility of nodes. 
The authors proposed a distributed anti-coordination game-based partially overlapping channels assignment (AC-POCA) scheme to minimize signal interference and maximize the communication capacity. 
In this game, the UAVs and devices are players and share the same channels. 
The strategies are the assignment of channels.
An $I-\text{Matrix}$ is a matrix used to record the interference of each user and determine whether the chosen channel is available to a given communication link.
Each player wants to be assigned a proper channel to maximize its throughput and minimize the interference from its neighbors.
Thus the utility of each player is a measure of the connectivity, which is $M_i$.
The total utility of the network is thus defined as $U_i(\Psi) = \sum_{i\in A}M_i$.
This utility function is found to be a potential function of the game and with the properties of a potential game, the authors were able to use the best response technique to obtain the Nash equilibrium.
The authors tested their algorithm on the mixed topology and dynamic topology scenario when the network topology keeps changing.
Simulation results demonstrate the impact of AC-POCA on convergence speed, signaling overhead, and throughput compared to the cooperative channel assignment game with best response and smoothed better response.
This algorithm proves to be very effective in a dynamic environment.

Reference \cite{UAV_offloading} tackled the offloading problem of the heavy computation tasks (e.g. pattern recognition and video reprocessing) to be completed by a fleet of UAVs. 
The problem was formulated as a non-cooperative game with $n$ players (i.e., the UAVs in the fleet) with  three pure strategies for each player. 
The three strategies are (1) perform their tasks locally, (2) offload them via a local wireless connection to a neighboring base station (BS), or (3) transfer through a cellular connection to an edge server (ES).
The utility is a linear combination of energy consumption, time delay, and computation cost, which is,
\begin{align}
 U = \alpha\sum_{i=1}^N T_i +\beta\sum_{i=1}^N E_i +\gamma \sum_{i=1}^N C_i,   
\end{align}
where $\alpha+\beta+\gamma = 1$, $N$ is the number of tasks, $T_i$, $E_i$, $C_i$ represent the time, energy overhead and communication cost respectively.
Thus, for UAV $i$, its utility function depends on which state it is on and has the form:
\begin{align}
&U_i(s_j,S_{-j}) =  \nonumber\\
&\left\{\begin{array}{lll}
U_{\mathrm{Local}} &= \alpha E_{\mathrm{Local}} + \beta T_{\mathrm{Local}}+\gamma C_{\mathrm{Local}}, &\mathrm{if}~ s_i = \mathrm{Local computing}\\ 
U_{\mathrm{Local}} &= \alpha E_{\mathrm{ES}} + \beta T_{\mathrm{ES}}+\gamma C_{ES}, &\mathrm{if}~ s_i = \mathrm{Offloading to ES}\\
U_{\mathrm{Local}} &= \alpha E_{\mathrm{BS}} + \beta T_{\mathrm{BS}}+\gamma C_{BS}, &\mathrm{if}~ s_i = \mathrm{Offloading to BS}\\ 
\end{array},\right.
\end{align}
where $E_{\mathrm{text}}$, $T_{\mathrm{text}}$, $C_{\mathrm{text}}$ are the energy consumption, time delay and computation cost for three actions, respectively.
This game is a potential game and the Nash equilibrium is found by a distributed offloading algorithm.
The simulation results indicate that this approach achieves in average of 19\%, 58\%, and 55\% better results compared with pure local computing, offloading to the edge server, and offloading to a base station respectively.
However, this algorithm faces a scaling problem if the network is very dense.

Finally, aerial UAVs face the challenge of malicious attacks such as jamming from aerial intruders.
Reference \cite{jamming} studied the jamming problem between an aerial jammer UAV and two communication UAVs.
The authors formulated this problem as a zero-sum pursuit-evasion game, in which the jammer UAV tries to maximize the jamming time, while the two communication UAVs aim to minimize the jamming time.
Then the \textit{Isaacs'} approach is used to derive the optimal control of each UAV, which turns out to be a bang-bang control verified by both theoretical analysis and simulation.
A drawback however is that each UAV needs to have complete knowledge of the state of the system.

Reference \cite{BayesianGame} utilized a Bayesian game for intrusion detection and ejection in a UAV-aided vehicular network.
A Bayesian game is a game in which each player only knows partial information about the payoff-relevant parameters, and the payoff is taken as the expectation over a distribution~\cite{Zamir2009}.
The motivation of this application is to provide a safety-oriented vehicular network by ejecting the suspected node so that important information can be exchanged among vehicles and UAVs.
The authors proposed two safety problems in UAV-aided communication systems. 
The first problem studies when the intrusion detection system should be activated, while the second problem focuses on the criterion to eliminate a seemingly malicious communication node. 
To solve these two problems, the authors modeled the attacks and defenses in an UAV system as two Bayesian games, where the information of the players is not known to each other. 
During the game, an intrusion detection node performs eight monitoring or waiting strategies, whereas a malicious node performs six strategies against UAV, cluster head or cluster members, either normal or malicious. Furthermore, both attackers and detectors can work in two modes. 
The attacker operates in a normal mode and an attacking mode, while the detector operates in a normal mode and a detect mode.
During the game, the attacker and detector gain a pre-defined profit with each strategy, which depends on the attacker's false positive rate and the detector's expected detection rate. 
It is shown in the paper that this Bayesian game has at least one Nash Equilibrium. 
At the equilibrium, the maximum profile $B$ gained by the attackers may be regarded as a threshold, which means that a normal node should perform malicious behaviors at a frequency less than $B$, but a malicious node performs bad behaviors more frequently than $B$. 
If such a node is found, then the intrusion detection system in the communication network should be activated, in order to find the attacker. 
The decision of the ejection, as is studied in problem two, follows a similar scheme. 
To decide whether a suspicious node should be cut off from the network, another Bayesian game is conducted. After the equilibrium is reached, the intrusion ejection system compares the rate of malicious behavior of a node with the profit at the equilibrium. 
If the former is larger than the later, then the node is probably performing attacks and should be ejected.
Simulation results demonstrate that the proposed framework exhibits a high detection rate and low false positive rate while requiring low communication overhead compared to existing frameworks.

\subsection{Stackelberg games}
A Stackelberg game is a hierarchical game comprised of two types of players: leaders and followers. 
In most cases, the leaders act first then the followers respond to the leaders' decisions. 
However, each leader must consider how the followers might respond to its decisions as well as to other leaders' decisions. 
A Stackelberg game is a common framework for analyzing resource allocation among consumers and provider companies.
More specifically, the companies decide the price of their resources and the consumers make decisions about the quantity they are going to purchase. 
The objective of both sides is to maximize their own benefits.
In wireless communications applications, Stackelberg games are used to study the pricing and bandwidth/power allocation problem when the two types of players (leaders, followers) are related and can have different game mechanisms.
For example, Reference \cite{stackelberg_pricing} studied the problem of downlink power allocation in a multi-UAV enabled wireless network by modeling it as a Stackelberg game.
In this game, the UAVs are the leaders choosing the optimal price to maximize their revenue defined as
\begin{align}
  \max \; U_{\text{UAV}}^j = \sum_{n=1}^{N}c_{jn}p_{jn}, \quad j \in \mathcal{M}, n \in \mathcal{N}_j  
\end{align}
where $c_{jn}$ is the price charged by the $j$th UAV to the $n$th user per unit power, $p_{jn}$ is the corresponding power, $\mathcal{M}$ denotes the set of UAVs, and $\mathcal{N}_j$ is the set of users served by the $j$th UAV.
The users are the followers selecting their optimal power strategy to maximize their revenue given by
\begin{align}
   \max \; U_{jn} = \log_2(1+\mathrm{SINR}_{jn}) - c_{jn} p_{jn}
\end{align}
with the constraint that $\sum_{n=1}^Np_{jn} \leq P_{\text{max}}$.
To make the game reach the equilibrium, a distributed iterative algorithm is proposed.
Simulation results also show that the proposed scheme performs better than the uniform power allocation scheme.

Likewise, Reference \cite{jointaccessselect} considered the UAV access selection and base station bandwidth allocation problems in a UAVs-assisted IoT network. 
In that case, the BSs are modeled as leaders and the UAVs are followers where the access competition among UAVs is formulated as a dynamic evolutionary game and the problem of bandwidth allocation of BSs is modeled as a non-cooperative game.

A Stackelberg game is also believed to be a promising formulation to the anti-jamming defence problem in wireless networks~\cite{Stackelberg_survey}.
A typical anti-jamming communication cycle includes three steps: jamming cognition, anti-jamming decision-making, and waveform reconfiguration.
Two common ways of addressing anti-jamming are power control and channel selection.
Stackelberg games were proposed in several works~\cite{stackelberg-bayesian,stackelberg-comm,stackelberg-IEEE} to solve the jamming power control problem in conventional communication networks. 
In these works, the legitimate users are the leaders and the jammer as the follower.
Both legitimate users and jammers need to choose their power to maximize their payoff based on SINR, throughput or transmission cost.

Anti-jamming power control in UAVs-assisted communication networks should consider the channel model of UAVs, the mutual interference, incomplete information constraint and the dynamic 3D flying environment. 
Reference \cite{SG_Antijamming_Bayesian} proposed a Bayesian Stackelberg game to model the competitive relations between multiple UAVs and a jammer.
To be more specific, the jammer acts as the leader while the UAVs are the followers.
The UAVs and jammers select their power control respectively to maximize their own payoff.
Note that incomplete information and observation errors have been considered for the UAVs.
The payoff of UAV $i$ is defined as follows:
\begin{align}
\resizebox{0.9\hsize}{!}{
    $U_i(P_i, P_{-i}, \Tilde{J}) =\sum_{g=1}^G \sigma_{\beta_i}(g)\log_2 \left( 1+\frac{\alpha_i P_i}{N_0+\beta_i(g)\Tilde{J} + \sum_{m\neq i}P_m\theta_{m,i}} \right)-C_u P_i$, 
}
\end{align}
where $J$, $P_i$ are the transmission power of jammer and UAV $i$ respectively, $\Tilde{J}$ is the observation value of $J$,  $\theta_{m,i}$ is the mutual interference gain which has $W$ states with probability $\sigma_{\theta_{m,i}}$, $\beta_i$ is the jamming gain which has $G$ states with probability distributions $\sigma_{\beta_i}(g)$, and $C_u$ is a constant.
The payoff of the jammer is 
\begin{align}
\resizebox{0.98\hsize}{!}{
    $V(J, P_1,...,P_N) = -\sum_{i,j,k}\sigma_{\alpha_i}(k)\sigma_{\theta_{m,i}}(w)\log_2\left(1+ \frac{\alpha_i(k)P_i}{N_0+\beta_i J+\sum_{m\neq i} P_m\theta_{m,i}(w)}\right)  -C_j J$ ,}
\end{align}
where $C_j$ is a constant, $\alpha_i$ is the transmission gain of UAV $i$ which has $K$ states with probability $\sigma_{\alpha_i}(k)$.
Then a sub-gradient-based Bayesian Stackelberg iterative algorithm is proposed to obtain the Stackelberg equilibrium, the existence and uniqueness of which are theoretically proven.
Simulation results illustrate the influence of incomplete information and observation errors.  They show for example, that if the observation error of the  jammer increases, the utility of UAV will decrease.
At the same time, the algorithm has a fast convergence rate and each player reaches its optimal transmission power within 5 iterations.
The main limitation of this work is that only one UAV jammer is considered.

\subsection{Mean field game}
The Mean Field Game (MFG) is a game-theoretic formulation suitable for dealing with a large number of agents.
MFGs approximate the interaction between one agent and other agents as that between the agent and the ``mean agent" of all others, which is commonly referred to as mean field approximation. 
The interaction of each individual player with the mean field effect of the rest of the population is generally captured through a Hamilton-Jacobi-Bellman (HJB) equation where the mean field function evolves following a Fokker-Planck-Kolmogorov (FPK) equation. 
The goal of each player is then simplified to maximize its own utility over a pre-defined period of time considering the collective behavior of the rest of the population.

Generally, MFGs are used when a large number of UAVs are involved. 
Indeed, the mean field approximation asymptotically achieves the $\epsilon$-Nash equilibrium of the original system when the number of agents goes to infinity~\cite{MFG_book_Prob}.
Researchers have used mean field games to model UAVs movement control problems in order to reduce energy consumption and maximize ground users coverage~\cite{MFG-movementcontrol,MFG_NN,MFG_NN_FL,AdaptiveCoverage}.

Reference \cite{MFG-movementcontrol} proposed a real-time MFG-based swarm movement control algorithm to minimize the weighted sum of each UAV's energy consumption per unit downlink rate and flocking cost.
In this way, both downlink transmission energy consumption and mechanical movement energy consumption are taken into account.
In this game, an individual UAV's velocity is determined by solving an HJB equation, and then the resultant UAV movements are obtained by solving a FPK equation in a windy environment. 
Each UAV can thereby decide its velocity using only its own location and channel states.
The dynamics of each UAV under windy environment is defined as
\begin{align}
    d z_i (t) = (v_i(t)+A) dt +\eta_A d W_i(t),
\end{align}
where $A$ is the average wind velocity, $\eta_A$ is the wind velocity variance, and $W_i$ is the standard Wiener process. 
The cost function of UAV $i$ is given by
\begin{align} \label{eqn:HJB}
    J_i(t) = \frac{1}{T}\int_{t}^{T}\omega_e E_i(v_i(t),z_i(t))+\omega_fF_i(v_i(t),z_i(t)) dt,
\end{align}
where $E_i(v_i(t),z_i(t))$ is the energy cost, $F_i(v_i(t),z_i(t))$ is the flocking cost, and $\omega_e$ and $\omega_f$ are the weighting factors.
Minimizing Equation \eqref{eqn:HJB}, a HJB equation is obtained.
Since in MFG, each agent is playing with the ``mean agent", the flocking cost can be written as follows:
\begin{align}
    F_i(v_i(t),z_i(t),m(z(t))) = \int_z \frac{m(z(t))\left \| v(z(t))-v_i(z_i(t)) \right \|^2}{(1/\gamma + \left \| z(t)-z_i(t) \right \|^2)^\beta}dz,
\end{align}
where $m(z(t))$ is the resultant UAV-position distribution.
This distribution is then the solution of a FPK equation which is coupled with the above HJB equation.
By solving the HJB-FPK equations, the optimal velocity is obtained.
The efficacy of this algorithm is verified by simulation using 3GPP air-to-ground channel model of UAVs.
The proposed algorithm saves up to 55\% average energy consumption per downlink rate compared to a baseline flocking scheme that does not consider energy efficiency under the same target collision probability.
Even though the solution to this problem is well-understood through the lens of mean field game formulation, it still incurs a large computational burden in solving these coupled partial differential equations (PDEs). 
In light of this difficulty, the authors of \cite{MFG_NN} utilized two separate neural networks (NNs) to approximate the solutions of HJB and FPK equations, thus providing one of the first links between game theory and machine learning.
Later in \cite{MFG_NN_FL}, the authors further combined federated learning with the neural network-based MFG method to help UAVs share parameters to achieve online path control and reduce computational burden. 

Reference \cite{AdaptiveCoverage} proposed a discrete-time MFG game framework where each UAV adjusts its velocity in order to increase the number of served users while simultaneously minimizing the flight energy consumption.
The aim of each UAV is also to optimize the velocity control (i.e., flight direction policy).
Unlike the above works, the flying model of UAV is assumed to be a discrete-time linear dynamic system and the UAVs are only allowed to fly in 9 directions (remain in place, move parallel to the coordinate axis, and move at a 45 degree angle with the axis of movement).
The cost function is defined as
\begin{align}
    J_i (u_i,m(t)) = \lim_{T\rightarrow \infty} E \sum_{t=0}^{T-1}(b\left \| x_i(t)-m(t) \right \|^2+u_i^T(t)Ru_i(t))
\end{align}
where $R$ is a pre-defined weighting matrix.
The optimal controller $u_i(t)$ is then obtained by solving this optimization problem analytically.

\subsection{Evolutionary game theory}
Evolutionary Game Theory (EGT) is a cross-field of evolutionary theory and game theory. 
The key idea behind EGT is the constitution of a population comprised of different phenotypes evolving over time. 
One important concept in EGT is that of Evolutionary Stable Strategies (ESS) defined as follows: 

\textbf{Definition (Evolutionary stable strategy~\cite{ESS_defi}):} Strategy $p^* \in S_n$ is evolutionary stable provided that for every other strategy $p \neq p^*$, there exists $\bar{\epsilon}(p) > 0$ such that the utility function satisfies
\begin{equation}
    U(p^*,\epsilon p+(1-\epsilon)p^*) > U(p,\epsilon p+(1-\epsilon)p^*),
\end{equation}
for every $0<\epsilon<\bar{\epsilon}(p)$.

EGT is used in wireless communication for access/mode selection and resource allocation when a population of players are involved. 
For example, Reference \cite{EGT-modeselec} proposed an EGT-based model selection approach in UAV-aided vehicular network.
In this application, three communication modes are available to the vehicles, namely, Vehicle to Base station (V2B), Vehicle to Vehicle (V2V), and Vehicle to UAV (V2U).
The vehicles need to decide which communication mode to choose in order to optimize the transmission reliability and the cost of resource utilization.
The payoff functions under three different choices are thus defined as 
\begin{align}
\left\{\begin{matrix}
\pi_{V2U} = k_u P_{UAV}(x)-q_u x_U \\ 
\pi_{V2B} = k_b P_{V2B}(x)-q_b x_B\\ 
\pi_{V2V} = k_v P_{V2V}(x)
\end{matrix}\right.,
\end{align}
where $P_{UAV}$, $P_{V2B}$, $P_{V2V}$ are respectively the transmission reliability of the three communication modes, $k_u$, $k_b$, $k_v$ , $q_u$, $q_b$ are all constants, $x_U$, $x_B$, $x_V$ are the proportions of players that choose the three strategies.
Usually, replicator dynamics (described in Equation~\eqref{eqn:replicator}) is used to describe the evolution process and capture the variation of the population state.
In this approach, each player decides to switch to another strategy if its profit is under the average payoff of the whole population. 
Thus, the replicator dynamics are given by
\begin{align}\label{eqn:replicator}
    \dot x_i = \sigma x_i(t)(\pi_i[x(t)]-\pi[x(t)]), \quad \forall \; i\in S,
\end{align}
where $i$ represents the strategy, $\sigma$ is a constant representing the speed of dynamic evolution, and $\pi[x(t)]$ is the average payoff of the whole population.
The authors then demonstrated the fast convergence of this evolutionary game based on replicator dynamics and higher transmission reliability with lower cost of resource utilization compared to the selfish and random selection schemes.

In \cite{jointaccessselect}, the authors studied the joint access selection and bandwidth allocation problem in an IoT system, where the access competition among groups of UAVs is formulated as a dynamic evolutionary game.
In this game, the players are all the UAVs and these UAVs decide which BS to connect with based on the BS's bandwidth and price.
If all players connect to the same BS, the bandwidth of this BS will be divided amongst them.
In this case, some players would rather connect to another BS to get a better payoff.
The payoff function is defined as
\begin{align}
    \pi_n^g(x) = \log\left (1+ \frac{k_n B_n R_n^g}{p_n N^g x_n^g} \right),
\end{align}
where $k_n$ is a predefined coefficient of the linear pricing function, $B_n$ is the allocated bandwidth of BS $n$, $p_n$ is the service price of BS $n$, $x_n^g$ denotes the proportion in group $g$ connecting to BS $n$, and $R_n^g$ measures the ergodic rate performance in group $g$ choosing BS $n$.   
This evolutionary game is solved using replicator dynamics and an ESS is obtained when the replicator dynamics reach an equilibrium.
Simulation results verify the fast convergence of this algorithm under different initial states.

\subsection{Summary and lessons learned}
We summarize in Table \ref{table:GT} several game theoretic formulations and their applications in UAVs-assisted wireless communication networks, covering the problems of task allocation, coverage maximizing, beaconing schedule, energy optimization, and so on.
Note that the ``drawbacks" term in the last column are a characteristic of a specific game in a specific situation, rather than an inherent weakness for all cases.

\begin{sidewaystable}[htbp]
\caption{Types of game theoretic approaches used in UAVs-assisted wireless communication networks.}
\begin{center}
\resizebox{\textwidth}{!}{
\begin{tabular}{|c|c|c|c|c|c|c|c|}
\hline
\textbf{Refs} &  \textbf{Description} & \textbf{Game model}  & \textbf{Players} &\textbf{Strategies} & \textbf{Utility} &  \textbf{Drawbacks}\\  \hline
\cite{coalition} &  Task allocation & \makecell{Hedonic coalition\\[-2pt] formation game} & UAVs, tasks& Form coalition &\makecell{A function of\\[-2pt] throughput, delay}  &
\makecell{NP-complete,\\[-2pt] sub-optimal}\\ \hline
\cite{Charging} & Recharging & Normal form game & UAVs &
\makecell{Probability of\\[-2pt] $R_{\text{OK}},R_{\text{NO}}$} & Residual energy & \makecell{matrix-based,\\[-2pt] information loss} \\ \hline
\cite{non-coop-coverage}& Positioning, coverage &
\makecell{Non-cooperative\\[-2pt] normal-form game}& UAVs& \makecell{Circle in current cell or \\[-2pt] move to adjacent cell} &  \makecell{Number of mobiles \\[-2pt] each UAV supports}& \makecell{High time complexity with \\[-2pt] the increasing size of players} \\ \hline
\cite{Koulali2016AGS} & Beaconing schedule & Sub-modular & UAVs& Beaconing period duration &\makecell{ Encounter rate,\\[-2pt] consumed energy }&  \makecell{Perfect rationality and \\ [-2pt]complete information}\\ \hline
\cite{POCA} & Channels assignment&\makecell{ Anti-coordination\\[-2pt] game} & UAVs, devices  &  Assignment of channel&\makecell{ Maximize the network\\[-2pt] throughput} & -\\ \hline
\cite{ UAV_offloading} & Offloading &  Non-cooperative  & Drones &\makecell{Local computing,\\[-2pt] offloading to ES,\\[-2pt] offloading to BS} & \makecell{Utility function\\[-2pt] that takes into account \\[-2pt] energy consumption, delay\\[-2pt] and communication cost} & Scalibility \\ \hline
\cite{jamming}& Jamming attack & \makecell{Zero-sum pursuit\\[-2pt] evasion game} & UAVs &Optimal control & Termination time &\makecell{ Complete knowledge\\[-2pt] of the state \\[-2pt] of the system}\\ \hline
\cite{BayesianGame}&  \makecell{Intrusion monitoring and\\[-2pt] attacker ejection }& Bayesian game &UAV and vehicles&\makecell{Monitor and eject\\[-2pt] malicious nodes}&\makecell{Protect communication\\[-2pt] network from attacks}&\makecell{Parameters are\\[-2pt] determined manually} \\ \hline
\cite{stackelberg_pricing}&Pricing and power allocation & Stackelberg game & UAVs, ground users & Power price, power & Revenue & - \\ \hline
\cite{jointaccessselect} & \makecell{Access selection,\\[-2pt] bandwidth allocation} & Stackelberg game & UAVs, BSs &\makecell{ Access selection,\\[-2pt] bandwidth allocation} & \makecell{UAVs (maximize payoff), \\[-2pt] BSs (maximize bandwidth allocation)} & - \\ \hline
\cite{SG_Antijamming_Bayesian}&\makecell{Anti-jamming \\[-2pt] power control} & \makecell{Bayesian \\[-2pt] Stackelberg game} & UAVs, jammer & Power control & \makecell{A function of\\[-2pt] throughput and\\[-2pt] transmission cost} &  Only one jammer considered\\ \hline
\cite{MFG-movementcontrol,MFG_NN,MFG_NN_FL} &  Minimize energy consumption & MFG & Massive UAVs& Optimal velocity& Energy consumption & High computation\\ \hline
\cite{AdaptiveCoverage} & Minimize energy consumption &MFG & UAVs  & Velocity control  & Energy consumption  & Ideal environment \\ \hline
\cite{EGT-modeselec} &  Mode selection & Evolutionary game  & Vehicles & \makecell{Selection of communication\\[-2pt] different communication modes}  &\makecell{ Transmission reliability\\[-2pt] and the cost of \\[-2pt]resource utilization} & Massive players\\ \hline
\cite{jointaccessselect} & Access selection& Evolutionary game &UAVs  & Connect to which BS & Payoff function of bandwidth and price& Massive players\\ \hline
\end{tabular}
}
\label{table:GT}
\end{center}
\end{sidewaystable}

The main lessons of this section include:
\begin{itemize}
  \item Game theory is a widely-used tool in the wireless communication field for modeling specific problems.
  \item The ultimate goal of a game is to find the (Nash) equilibrium.
  \item Different game types are appropriate for different problems in UAVs-assisted communication networks.
    \item The time complexity of conventional game theory solutions grows with the increase in the number of players.
  \item Mean field games and evolutionary games are potentially useful in massive UAVs network communication problems.
\end{itemize}
\section{Machine learning in UAVs-assisted wireless communication networks} \label{sec:ML}
Machine learning techniques were introduced into the wireless communication field due to their ability to predict future network states, generalize to new unseen network states, and scale to large-size networks~\cite{security}.
Machine learning methods are generally divided into supervised learning, unsupervised learning, and reinforcement learning methods.
With the improvements in parallel computing and graphics processing units (GPU), neural networks (NN) became a powerful tool for machine learning. 
Notable structures of NN include deep feed-forward networks (DFF), convolutional neural networks (CNN), and recurrent neural networks (RNN). 

ML tools have been applied in the U-WCN arena for modelling, predicting and monitoring traffic patterns~\cite{RLCoverage,DRLCoverage,ML-deployment}, device locations, network access and rate control~\cite{LSTM_app,CacheESN}, connectivity preservation, resource allocation and interference management~\cite{QL,DDPG-RA}. 
These applications have benefited from recent advances in both theory and computational tools such as Tensorflow, Pytorch, and MATLAB's machine learning toolbox.

\subsection{Neural networks} 
Powerful ML techniques such as deep learning and reinforcement learning are now used in the UAVs-assisted wireless communication field \cite{ML_UAV_intro,security,ML-UAV-survey,ML-deployment}. 
Compared to conventional model-based approaches, ML tools allow designers to take into account application-specific issues, such as the type of UAVs, Doppler effects, cache management, dynamic positioning, interference management, and load balancing \cite{ML_UAV_intro}. 

In \cite{ML-deployment}, the authors proposed an ML framework based on a Gaussian Mixture Model (GMM) and Weighted Expectation Maximization (WEM) algorithm to predict potential network congestion.
Based on the predicted traffic, the optimal deployment of UAVs is then obtained by minimizing the transmission power and mobility powers. 
In that work, the authors used the actual dataset of a Chinese City Cellular Traffic Map. The dataset is composed of the number of aerial users that are offloaded from a BS at location $(x,y)$ to a UAV during time interval $[t,t+T]$, and the amount of cellular traffic that a UAV needs to provide for the aerial users from a BS at $(x,y)$ at $[t,t+T]$.
The aim is to predict the total number of aerial users, the spatial distribution of aerial users, and the spatial distribution of aerial data traffic in a geographical area 
$\mathcal{A}$.
Using GMM and WEM, the authors were able to predict the cellular traffic allowing for a constrained optimal problem to be solved in order to minimize the total power for downlink transmission and mobility.
The simulation results show that the proposed algorithm reduces the power consumption for downlink transmission and mobility by over 20\% and 80\% respectively compared to a more traditional optimization approach without machine learning used.
In the following subsections, different deep neural networks (e.g., convolutional neural networks, recurrent neural networks, spiking neural networks) and their applications in UAVs-assisted wireless communication networks will be reviewed.
\subsubsection{Convolutional neural networks}
Convolutional Neural Networks (CNN) were initially proposed and used in computer vision. 
A CNN consists of an input layer, several hidden layers and an output layer. 
The name ``convolutional" originated from the use of the convolution operator. 
The hidden layer of a CNN contains a convolutional layer, an activation layer, a pooling layer, and a fully connected layer.  CNNs are useful because of their image processing ability, which can provide UAVs with vision-based sensing capabilities. 
By combining with reinforcement learning algorithms or recurrent neural networks, CNNs are playing an increasing role in UAVs-assisted wireless communication networks. 
For example, in a cellular-UAV network, CNNs help the UAVs identify the location of ground BSs, ground user equipment, and other UAVs in the network. 
Such information can then be fed into a recurrent neural network to help individual UAVs make decisions about their future movement in order to minimize the interference and latency at each time instant~\cite{security}. 
Another potential application of CNNs lies in UAV-enabled edge caching,  where a CNN extracts and stores common features of the data files (videos, images, etc.) requested by different users, then uses these features to predict a user's video requests and preference~\cite{security}.
\subsubsection{Recurrent neural networks}
\begin{figure*}[!htb]
  \includegraphics[width=\linewidth]{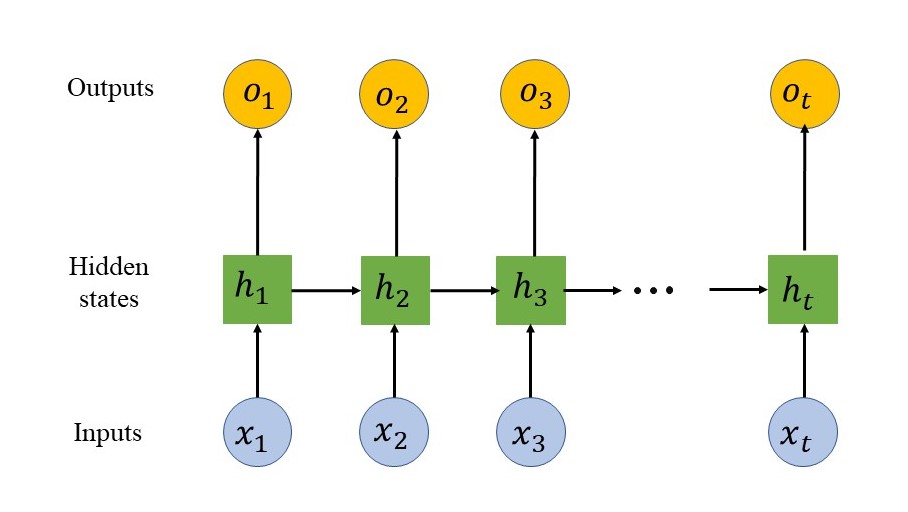}
  \caption{RNN structure.}\label{fig:RNNstructure}
\end{figure*}

Recurrent Neural Networks (RNN) are a class of artificial neural networks that make use of sequential information. 
Fig.~\ref{fig:RNNstructure} presents the illustration of an RNN structure. 
Such a structure is able to capture long-term dependencies hidden in the dataset. 

Echo state networks and long-short-term memory networks are two widely-used RNN structures.  
The Echo State Network (ESN) is a practical type of recurrent neural network with a sparsely connected hidden layer.  
ESN is characterized by its adaptive memory, which enables it to store previous state information in order to predict future states of UAVs.  
Reference~\cite{CacheESN} studied the problem of proactive deployment of cache-enabled UAVs for optimizing the Quality-of-Experience (QoE) of wireless devices in a cloud radio access network.
In this model, a conceptor-based ESN is deployed to predict the content request distribution and mobility pattern of each user, leveraging the users' visited locations, requested contents, and other human-centric information.  
Then, with these predictions, the authors tried to find the user-UAV associations, the optimal UAVs' locations, and the contents of the cache at UAVs by formulating an optimization problem to maximize the users' QoE while minimizing the UAVs' transmission power.
The dataset is from BUPT and Youku recording the real pedestrian mobility patterns and content transmission.
Simulation results show that the proposed algorithm achieves 33.3\% and 59.6\% gains in terms of the average transmit power and the percentage of the users with satisfied QoE compared to that of a benchmark algorithm without caching and a benchmark solution without UAVs.
The advantage of using ESN here is that the users' mobility pattern and content request distribution have some time-dependent and spatial statistical characteristics.

A Long-Short-Term Memory (LSTM) network is another specific type of recurrent neural network that can learn long-term dependencies~\cite{LSTM-ori}.
LSTM networks have been successfully used in classification, image recognition, and machine translation fields~\cite{sentimentclassfication,LSTMgesturerecog,LSTMtranslation}. 
With three gated units, an LSTM network solves the gradient diminishing problem of traditional RNN structures.

Recently, researchers proposed integrating LSTM into D2D communication systems.
For example, \cite{DL_UAV} designed an integrated LSTM and Multi-Layer Perceptron (MLP) architecture to determine the position of a UAV in order to maximize the A2G link access coverage performance, while minimizing the transmission power and maximizing the user's throughput. 
In this experiment, the authors considered three UAVs connected through wireless multi-hop backhauls to the core network. 
To collect data, the authors designed the data acquisition procedures and environment at the National Taipei University of Technology with 900 MHz band.
Data of the A2G link access coverage probability, Line of Sight/Non-Line of Sight (LoS/NLoS), elevation angle, Received Signal Strength (RSS), Signal-to-Noise-Ratio (SNR), and user-to-user distance are collected.
The target area is divided into grid points and the data collected at 722 reference points are used as training samples, while data collected at another 85 reference points are used for testing.
The collected data is then sent to the MLP-LSTM neural network structure as the input.
Using the proposed MLP-LSTM structure, the algorithm finds the UAV position that maximizes the throughput.
The authors compared this MLP-LSTM scheme performance with Support Vector Machines (SVM), LSTM, and MLP algorithms in three scenarios: using the original datasets; using reduced features only and estimating the values of user throughput for each user at each grid point; and using reduced data collected on different days/times and finding the grid points in which users achieved maximum and total throughput.
The experiments indicate that the UAV positioning provides an accuracy level of 94.73\%, 98.33\%, and 99.53\% respectively in three scenarios and outperform SVM, MLP, and LSTM.

Reference~\cite{LSTM_app} proposed the use of LSTM to predict the classification of potential content providers so that the D2D communication system between the content provider and the content requester achieves a desired level of confidentiality.
In that article, LSTM selects the optimal D2D transmitter for the content requester based on experience and real-time information of the content requester, such as the amount of content requested, the mobile status of the content carriers, the distance between the content carriers and the content requesters, and the remaining energy of the UAV flying base station. 
Using simulation, the authors showed that the LSTM scheme improves the security capabilities of the system compared to the random-based scheme.
\subsubsection{Spiking neural networks}
Spiking Neural Networks (SNN) are novel artificial neural networks that mimic the operation of brain neurons. 
Liquid State Machine (LSM) is a particular type of SNN with five components: agents, input, output, liquid model, and output function. 
LSM is proposed to handle continuous-time inputs and to compute at various time scales. 
It has two advantages over traditional artificial neural networks, namely, fast real-time decoding of signals and high information carriage capacity by adding a temporal dimension ~\cite{ANN-tutorial}, and has been used for optimizing resource allocation in wireless communication with UAVs \cite{SNN-LSM}.

Reference~\cite{SNN-LSM} proposed a distributed algorithm based on LSM to jointly optimize the user association, spectrum allocation, and content caching.
The LSM stores the users' behavior information and tracks the state of the network over time in order to predict the content request distribution, and automatically adapts spectrum allocation to the change of the network states.
In this algorithm, a cloud first predicts the content request distribution of each user using an LSM-based approach.
Then with this distribution, each UAV finds the optimal user association by using an $\epsilon$-greedy mechanism.
In this way, this algorithm solves the challenge of the original problem, which is a nonlinear discrete optimization problem.
Simulation results show that it outperforms the Q-learning algorithm (introduced in Section~\ref{subsec:Ql}) in terms of the average number of stable queue users.

The machine learning algorithms described so far, require that all data are sent to a central location.
To address this shortcoming, federated learning emerged as an effective tool to implement machine learning in a distributed fashion.
\subsection{Federated learning}
Federated Learning (FL) is a concept proposed by Google researchers~\cite{googleFL2}.
It involves training the model in a central server, while keeping the data localized, thus realizing the goal of preserving privacy and safety. 
\textbf{Algorithm \ref{algo:FDL}} summarizes the FDL algorithm presented in~\cite{UAV_FDL}.
In this algorithm, $N$ UAVs store their own data and train a separate model on the data.
Then these model parameters are aggregated by averaging to obtain a final model.
With this mechanism, on one hand, a loss of one UAV's data will not greatly affect the whole system performance.
On the other hand, storing the data in each UAV can reduce the energy loss of transmitting all the data to a central controller and protect privacy ~\cite{FL-wireless,dynamicFL,incentiveFL}. 

Reference \cite{FL-wireless} formulated federated learning over a wireless network as an optimization problem thus providing insight into the compromise between energy consumption, learning accuracy, and time. 
Reference \cite{incentiveFL} adopted contract theory to design an effective incentive mechanism to stimulate the mobile users with high-quality data to participate in federated learning in order to solve the heterogeneity problem.
Because UAVs are resource-constrained devices while traditional ML-assisted schemes require UAVs' data to be sent and stored in a centralized server, distributed ML is needed in the UAVs-assisted wireless communication setting. 

Reference \cite{UAV_FDL} first introduced federated deep learning (FDL) concepts for UAV-enabled wireless applications and the authors discussed the key technical challenges, open issues, and future directions on FDL-based approaches. 
Basically, the FDL training process of UAV-based networks comprises three steps.
The first is the training initialization. 
A server specifies the required data type and training hyper-parameters, together with an initial global model $G_0$ and broadcasts them to the UAVs.
The second step is the UAVs' model training process.
Each UAV collects data, keeps the data to itself, and updates parameters of its local model $L_i^j$.
Then the updated parameters are sent to the server.
The final step is the global model aggregation.
The server aggregates these local models and sends back the updated model parameters to the UAVs.
Recently, researchers have started to examine decentralized federated learning, to eliminate the need of a centralized server.
Such works can be found in ~\citep{lalitha2019peer,savazzi2020federated,taya2021decentralized}, which provide fully decentralized framework for localized data and have greater potential in future IoT applications.
Despite the above advantages, FDL still faces challenges from heterogeneous data distributions in real applications, and the lack of theoretical guarantees of convergence.  Other problems will arise if FDL is applied to UAVs-assisted wireless communication given that UAVs are operating in a highly-dynamic environment.

\begin{algorithm}[!htbp] \label{algo:FDL}
 \KwData{Number of UAVs $N$, number of local epochs $E$, batch size $B$, learning rate $\eta$, number of server rounds $R$}
 initial global model $G_0$\;
 \For{$j=1$ to $R$}{
  $P=$ random set of UAVs of $N$\;
  \For{each UAV $i$ in $P$ in parallel }{$L_i^{j+1} \leftarrow \mathbf{ClientUpdate}(i, L^j)$\;}
    $G^{j+1}\leftarrow \frac{1}{|P|}\sum_{i=1}^{|P|}L_{i}^{j+1}$\;
 }
 \Return{$G^{j+1}$.}
 
 \textbf{ClientUpdate($i$,$L$)}:
 
 \For{$e=1$ to $E$}{
 batches $\leftarrow$ split dataset into batches of size $B$\;
 \For{each batch $b$}{
 $L \leftarrow L-\eta \bigtriangledown f(L,b)$\;}}
 \Return{$L$ to UAV.}
 \caption{FDL for FL server}
 
\end{algorithm}
\subsection{Reinforcement learning}
Reinforcement Learning (RL) is a sub-field of machine learning.
Detailed introductions and examples of reinforcement learning may be found in \cite{RL_sutton}. 
There are four main elements for an agent in a reinforcement learning system: a policy, a reward, a value function, and a model of the environment. 
Compared with supervised and unsupervised learning methods, RL-based algorithms have the advantage of learning in an unknown environment with a pre-designed reward. 
In particular, RL algorithms are used in UAVs-assisted wireless communication services to solve deployment, resource allocation, navigation, and control problems. 
In the following subsections, two commonly-used reinforcement learning algorithms, Q-learning and deep deterministic policy gradient, as well as their corresponding applications are reviewed.
\begin{figure}[htbp] 
\centerline{\includegraphics[width=1.3\textwidth]{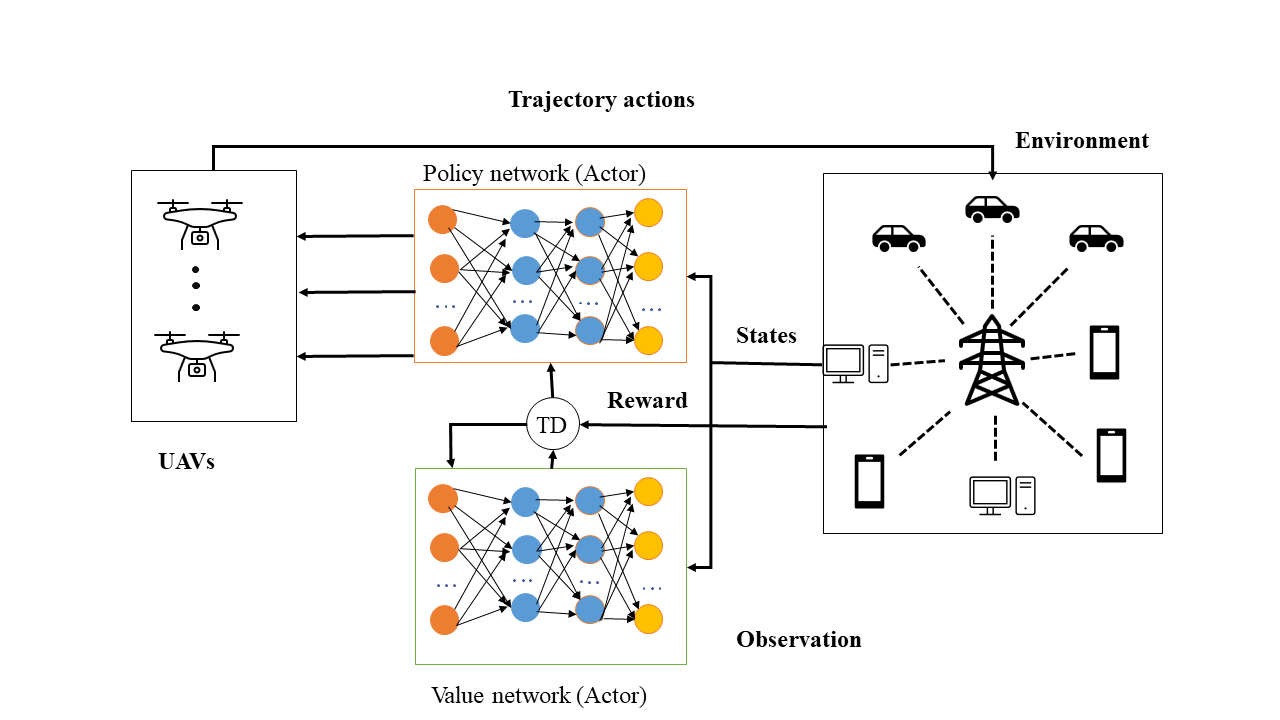}}
\caption{DDPG-based UAV trajectory planning.}
\label{figure:RL-AC}
\end{figure}
\subsubsection{Q-learning} \label{subsec:Ql}
Q-Learning (QL) is a model-free reinforcement learning algorithm that guides an agent to take a specific action in a given environment. 
QL provides an optimal action selection policy for a given finite Markov decision process \cite{QL_intro}. 
A typical Q-learning algorithm is shown in \textbf{Algorithm~\ref{algo:QL}}.
To alleviate the space complexity of search on the Q-table, the deep Q-network, which uses a neural network to map the input states to the action value, was proposed.
Q-learning was introduced to the study of UAVs-assisted wireless communication networks in order to solve the trajectory planning, 3D deployment, security, and  resource allocation problems.

\begin{algorithm} \label{algo:QL}
 initial $Q_0$; discount factor $\gamma$; learning rate $\alpha$\;
 \For{$t$ in epoch}{
 At time $t$ with state $s_t$, selects an action $a_t$, observe a reward $r_t$, obtain next state $s_{t+1}$\;
 Update Q table:
  \begin{align} \label{eq:QL}
     Q^{new}(s_t,a_t)\leftarrow Q(s_t,a_t)+\alpha(r_t +\gamma \mathrm{max}_{a} Q(s_{t+1},a)-Q(s_t,a_t)).
 \end{align}
 }
 \caption{Q-Learning}
\end{algorithm}

As an example, Reference \cite{QL-emergency} proposed a Q-learning algorithm to find the best 3D positioning of multiple drone small cells in an emergency scenario. 
The main goal of this work is to maximize the number of users served by the drones with the constraints of both backhaul and radio access network. 
The state space is the position of UAV, the action space is \{ Up, Down, Left, Right, Forward, Backward, Keep still\}, and the reward is the total number of users allocated to the UAV.
Then, $\epsilon$-policy is used to find the optimal solution.
The proposed algorithm is shown to be robust to different network conditions, such as the position of other drones, interference between drones, as well as user movements and their constraints.
Simulation results also show that the proposed algorithm has advantages over random position, fixed position, and circular position schemes in terms of two measures: users' throughput dissatisfaction and the percentage of users in outage.
Similarly, in \cite{QL-trajectory}, one UAV was chosen as a base station in order to provide network services to multiple users. 
The main goal of that work is to optimize the trajectory of the UAV in order to maximize the sum rate of transmission (i.e., reward) during flying time. 
In such a problem, the state-space is composed of the position of the UAV while the action space contains \textit{\{up, right, down, left\}} on the same plane. 
The authors compared the table-based Q-learning and NN-approximator-based Q-learning approaches and showed that both converge to the desired trajectory. 
Reference \cite{zhanghanbook} presented Q-learning for a UAV trajectory design problem. 
In this problem, the finite state-space contains all possible locations of the UAVs, and the algorithm selects from a corresponding finite set of actions (27 directions).
The reward of a UAV is designed as the total number of
successful valid sensory data transmissions for its task.
Then Q-learning algorithm is used to find the best action of each UAV.
Even though this single-agent Q-learning has many favorable properties due to its small state-space and action sets, it does not account for the states and strategies of other UAVs. 
To solve this problem, in the same book, the authors presented a multi-agent Q-learning algorithm called opponent modeling Q-learning. 
This multi-agent reinforcement learning can better model the cooperation or competition relations among agents.
However, the challenge of multi-agent reinforcement learning is that convergence can only be guaranteed under restrictive assumptions.
Moreover, the above three works assume that UAV can only move in horizontal directions, which is limiting in real applications.

Reference~\cite{Attack-PT-Q} applied prospect theory to formulate a subjective smart attack game for the UAV transmission. 
In this game, an attacker UAV can choose from three attack types (jamming, spoofing, and eavesdropping), and the defender UAV chooses the transmit power ($B$) on multiple radio channels to resist the smart attack. 
The prospect theory-based utility function of the defence UAV is defined as

\hspace{-20pt}
\resizebox{\linewidth}{!}{\parbox{1.1\linewidth}{
\begin{numcases}{U(x,y)=}
\sum_{i=1}^B\left(h_{T,i}^{(k)}-h_{E,i}^{(k)}\right)x_i - \mu\sum_{i=1}^Bx_i, & if  $y=-1$ \nonumber \\ 
\sum_{i=1}^B h_{T,i}^{(k)}x_i-\frac{C_m}{L}\sum_{l=0}^Llw_A(\beta_l)-\mu\sum_{i=1}^Bx_i, & if $y=-2$ \\ 
\sum_{i=1}^B h_{T,i}^{(k)}x_i-\mu\sum_{i=1}^Bx_i -\frac{1}{L}\sum_{l=0}^L lw_A(\eta_l)\sum_{i=1}^B\frac{h_{T,i}^{(k)}h_{J,i}^{(k)}x_iy_i}{\sigma+h_{J,i}^{(k)}y_i}, & if  $y \geq 0$ \nonumber
\end{numcases}}}
where $x=\{x_i\}$, $y=\{y_j\}$ are the strategy set of defence UAV and attacker respectively, $h_{T,i}^{(k)}$ is the channel power gain of defence UAV and its user, $h_{E,i}^{(k)}$ is the wiretap channel gain, $h_{J,i}^{(k)}$ is the jamming channel gain, and $\omega_A(p)$ is the subjective probability viewed by defence UAV.
Deep Q-learning algorithms (i.e., DQN) are then developed to achieve optimal power allocation against smart attacks.
Simulation results reveal that DQN-based strategy has the highest safe rate, secrecy capacity, and SINR compared to pure Q-learning-based strategy and WoLF-PHC (Win or Learn Faster-Policy Hill Climbing)-based strategy.
However, this performance comes at the cost of highest computational complexity and DQN takes a much longer time to make a decision.

In \cite{QL}, the authors investigated the dynamic resource allocation of multiple UAVs within a Multi-Agent Reinforcement Learning (MARL) framework.
The goal for each UAV $m$ is to jointly select the user ($a_m$), power level ($p_m$), and sub-channel ($c_m$) to ensure that the SINR provided by the UAVs is greater than a given threshold. 
The state of UAV $m$ at time $t$ is defined as
\begin{align}
s_m(t) = \left\{\begin{matrix}
1,\quad  \gamma_m(t)\geq \bar{\gamma}\\ 
0, \quad \gamma_m(t)< \bar{\gamma}
\end{matrix}\right.,
\end{align}
where $\bar{\gamma}$ is the threshold of satisfactory SINR.
The reward function is 
\begin{align}
R_m(t) = \left\{\begin{matrix}
\frac{W}{K}\log_2(1+\gamma_m(t))-\omega_m P_m(t), \quad \text{if} \;\gamma_m(t) \geq \bar{\gamma}_m\\ 
0, \text{else}
\end{matrix}\right.,
\end{align}
where $\gamma_m$ is the observed SINR of UAV $m$, $\omega_m$ is the cost per unit level of power, $P_m(t)$ is the transmit power of UAV $m$ at time slot $t$, and $\frac{W}{K}$ is the sub-channel bandwidth.
Each UAV runs its decision algorithm independently, but all share a common structure based on Q-learning. 
The efficacy of the proposed MARL framework is shown via simulation and has a higher average reward compared to matching theory-based resource allocation and random user selection algorithms.

The above works are all rooted in the offline Q-learning framework, which suffers from the well-known curse of dimensionality when the state and action spaces are large.

Considering this drawback, \cite{DQL_power} proposed an on-board (or online) deep Q-learning technique to minimize the overall data packet loss of sensing devices. 
In this problem, the battery levels of ground devices, the queue lengths of the ground devices, the channel quality between the UAV and the device, and the location of the UAV are defined as the state.
The selection of ground devices, the modulation of the device, and the instantaneous patrolling velocity of the UAV are the actions.
Then a deep Q-network algorithm is used to learn and decide the device to be charged and interrogated for data collection and the instantaneous velocity of the UAV.
This on-board deep Q-network has two separate Q-networks with current weights and old weights.
Simulation results indicate that this algorithm has lower network costs and packet loss rates compared to other on-board scheduling policies.

Traditional Q-learning uses the same values both for selecting and evaluating an action, thus suffering from the overestimation of action values under certain conditions. 
Double Q-learning was proposed to solve this problem \cite{double_Qlearning}.
For double Q-learning, the selection and evaluation of an action are decoupled by using two value functions.
The two value functions are learned by assigning each experience randomly to update one of them with weights $\theta$ and $\theta'$ respectively.
During each update, one set of weights is used to determine the greedy policy and the other to determine its value.
The clear difference between Q-learning and double Q-learning can be displayed by the following equations \cite{double_Qlearning}:
\begin{align}
\left\{\begin{matrix}
Y_t^\mathrm{Q} = R_{t+1}+\gamma Q (S_{t+1},\argmax_{a} Q(S_{t+1},a;\theta_t);\theta_t)\\
Y_t^{\mathrm{DoubleQ}} = R_{t+1}+\gamma Q (S_{t+1},\argmax_{a} Q(S_{t+1},a;\theta_t);\theta'_t)
\end{matrix}\right.,
\end{align}
where $Y_t^{\mathrm{Q}}$ and $Y_t^{\mathrm{DoubleQ}}$ are the target value of Q-learning and double Q-learning respectively.

One recent application of double Q-learning is in \cite{doubleQ_application}. 
In this article, the authors proposed an on-board double Q-learning scheduling algorithm for a UAV to select the IoT node for data collection and microwave power transfer along a predetermined flight trajectory.
Similar to \citep{DQL_power}, the objective is to minimize data packet loss resulting from buffer overflow and channel fading.
The action space is the selection of IoT nodes while the state space contains the battery levels, queue length of the IoT nodes, and the channel conditions between the IoT nodes and UAV.
Then a double Q-learning algorithm is used to find the best selection of IoT nodes to reduce packet loss.
To verify its efficacy, the authors compared their algorithm with the Q-learning algorithm.
Simulation results show that double Q-learning outperforms Q-learning in both packet loss rate and learning error.
Similarly, the proposed algorithm has an advantage over two other scheduling algorithms, which are called ``Longest Queue Scheduling Algorithm" and ``Longest Queue Lowest Battery algorithm".

\subsubsection{Deep deterministic policy gradient}
Deep Deterministic Policy Gradient (DDPG) is a model-free, off-policy actor-critic algorithm that concurrently learns the Q function and policy with neural networks.
Both the critic and actor networks are parameterized using neural networks. 
DDPG learns policies in high-dimensional, continuous action spaces \cite{DDPG}. 
DDPG and its variants have been studied in robotics~\cite{DDPG_robotics}, self-driving~\cite{wang2019deep}, physical control domains and games such as Atari, chess, and others~\cite{ddpg_Atari}. 
A typical DDPG algorithm \cite{DDPG} is shown in \textbf{Algorithm \ref{algo:DDPG}}.

\begin{algorithm} \label{algo:DDPG}
 Randomly initialize critic network $Q(s,a|\theta^Q)$ and actor $\mu(s|\theta^\mu)$ with weights $\theta^Q$ and $\theta^\mu$\;
 Initialize target network $Q'$ and $\mu'$ with weights $\theta^{Q'} \leftarrow \theta^Q$, $\theta^{\mu'} \leftarrow \theta^\mu$\;
 Initialize relay buffer $R$\;
 \For{episode =1, M}{
 Initialize a random process $\mathcal{N}$ for action exploration\; 
 Receive initial observation state $s_1$\;
 \For{t=1,T}{
 Select action $a_t = \mu(s_t|\theta^\mu) + \mathcal{N}_t$ according to the current policy and exploration noise;\
 Execute action $a_t$ and observe reward $r_t$ and observe new state $s_{t+1}$\; 
 Store transition $(s_t,a_t,r_t,s_{t+1})$ in $R$\;
Sample a random mini batch of $N$ transitions $(s_i,a_i,r_i,s_{i+1})$ from $R$\; 
Set $y_i = r_i+\gamma Q'(s_{i+1}, \mu'(s_{i+1}|\theta^{\mu'})|\theta^{Q'})$\;
Update critic by minimizing the loss:~$L=\frac{1}{N}\sum_i(y_i-Q(s_i,a_i|\theta^Q))^2$\;
Update the actor policy using the sampled policy gradient:\

\quad \quad$\bigtriangledown_{\theta^\mu}J\approx \frac{1}{N}\sum_i\bigtriangledown_a Q(s,a|\theta^Q)|_{s=s_i,a=\mu(s_i)}\bigtriangledown_{\theta^\mu}\mu(s|\theta^\mu)|_{s_i}$\;
  
Update the target network: \\
\quad \quad $\theta^{Q'} \leftarrow \tau\theta^Q +(1-\tau \theta^{Q'})$\;
\quad \quad $\theta^{\mu'} \leftarrow \tau\theta^\mu +(1-\tau \theta^{\mu'})$}.
}
\caption{DDPG Algorithm}
\end{algorithm}

DDPG extends the scope of Q-learning and has advantages over Q-learning in dealing with a continuous action space and high-dimensional problems.  
It is thus used in UAVs-assisted wireless communication problems to help solve the trajectory design, resource allocation, and deployment problems~\cite{RLCoverage,DRLCoverage,DDPG-RA}. 

Reference \cite{RLCoverage} proposed a DDPG-based algorithm for learning the optimal trajectories of a swarm of UAVs to efficiently maximize their coverage for vehicles on highways with poor cellular infrastructure and highly-dynamic environment. 
Fig.~\ref{figure:RL-AC} shows this application scenario where DDPG is used in a dynamic UAV-vehicular environment to optimize UAVs' trajectory. 
In this article, each UAV carries out a continuous control task to serve the vehicles on a highway.
The inputs of the UAVs in the dynamic vehicular environment at time slot $n$ include: the remaining energy of each UAV, the number of vehicles residing within the considered highway segment, the instantaneous positions of vehicles, ground level position of each UAV, the status of the UAVs describing whether a UAV is deployed or not, and the coverage indicators of each vehicle.
Each UAV takes an action which gives a traveling distance in a specific direction.
The reward takes into consideration of several quantities: the coverage penalty due to non-coverage, 
the deployment penalty due to the deployment of a new UAV, the energy penalty due to traveling, and the penalty if the UAV flies outside the given segment.
By using an actor-critic algorithm, the UAVs learn their flying trajectory and achieve an effective coverage with a minimum number of UAVs.
The proposed algorithm is compared with three other approaches (namely, random UAV dispatching approach, fixed dispatching rate approach, and fixed hovering UAVs approach).
It is shown that the proposed algorithm improves upon all three algorithms in terms of the number of required UAVs since it allows a UAV to dynamically predict and adapt its trajectory.
The proposed algorithm also achieves the same coverage with less energy consumption.
The drawback of this algorithm, however, is that it takes a long time (16 hours) to learn in the vehicular environment to obtain a good performance.

With the same aim, Reference~\cite{DRLCoverage} proposed a DDPG-based method to find a flying control policy for UAVs, that will jointly maximize coverage and fairness while minimizing energy consumption.
As opposed to \cite{RLCoverage}, the state-space contains three quantities: the current coverage score of each Point-of-Interest (PoI), the current coverage state of each PoI, and the current energy consumption.
The reward function is given by
\begin{align}
    r_t = \frac{f_t(\sum_{k=1}^K \Delta c_k^t)}{\sum_{i=1}^N\Delta e_i^t},
\end{align}
where $f_t$ is the fairness index, $\Delta c_k^t$ is the increment coverage score, and $\Delta e_i^t$ is the incremental energy consumption.
The proposed algorithm learns the UAVs' flying distance and flying direction.
Simulation results show that the proposed algorithm outperforms two baselines (i.e. Random and Greedy policy) in terms of average coverage score and average energy consumption in spite of the number of UAVs used and the coverage range.

By the same token, online DDPG is needed for future U-WCNs~\cite{DDPG-RA}.
To jointly optimize the flight control of the UAV and data collection scheduling along the trajectory in real time, \cite{DDPG-RA} proposed a new online flight resource allocation scheme based on a DDPG algorithm.
In particular, the flight resource allocation problem is formulated as a Markov decision process, where the network states consist of the battery level, data queue length, signal-to-noise-ratio of the channel, and the location of the UAV.
The action set is composed of the heading, the patrol velocity of the UAVs and ground nodes selection for data collection.
The heading and patrol velocity are in continuous action spaces and the reward is the packet loss of the network.
Simulation results show the convergence of this algorithm.
Furthermore, the same problem was extended and studied in \cite{DDPG_MC} by considering the real flying model and channel model of UAVs.
Then an on-board DDPG-based maneuver control was proposed to jointly optimize the online maneuver control and communication schedule.
 
\subsection{Summary and lessons learned}
This section presented several popular machine learning algorithms and their applications to various problems in UAVs-assisted wireless communication networks. 
The above discussed machine learning frameworks and their corresponding applications in UAVs-assisted wireless communication networks are summarized in Table \ref{table:ML}.
\begin{sidewaystable}[htbp]
\caption{Types of machine learning approaches used in UAV-assisted wireless communication networks.}

\begin{center}
\resizebox{\textwidth}{!}{
\begin{tabular}{|c|c|c|c|c|}
\hline
\textbf{Refs}  & \textbf{Description} & \textbf{ML algorithms} & Dataset/State & Outputs/Action\\  \hline
\cite{ML-deployment}& On-demand deployment of UAV & GMM & \makecell{Number of users offloaded, \\[-2pt] intensity of cellular traffic} & \makecell{Number of aerial users, spacial distributions \\[-2pt] of  aerial users and data traffic} \\ \hline
\cite{CacheESN} & Proactive deployment& ESN & \makecell{Users' visited locations, \\[-2pt] contents requested, etc.}  & Request distribution, mobility pattern \\ \hline
\cite{DL_UAV} & UAV positioning& LSTM, MLP& LOS, elevation angle, etc.& Position, throughput\\ \hline
\cite{LSTM_app}& Security &LSTM & Information of the content requester & Optimal D2D transmitter\\ \hline
\cite{SNN-LSM}& Resource allocation & LSM & Users' information, UAVs' action &  Content request distribution, UAVs' action\\ \hline
\cite{UAV_FDL}  & Mechanisms, challenges & Deep federated learning  & -&-\\ \hline
\cite{zhanghanbook,QL-emergency,QL-trajectory} &Positioning, trajectory planning&  Q-learning & Position of UAVs & Up, down, left, etc.\\ \hline
\cite{QL}  & Resource allocation & Multi-agent Q-learning & 1, 0& \makecell{Selection of communicating user,\\[-2pt] power level and sub-channel}\\ \hline
\citep{Attack-PT-Q} & Security & DQN & Attack mode & Transmit power selection \\ \hline
\cite{DQL_power} & Trajectory planning, power transfer & Online Q-learning & Battery level, queue length, etc. & Selection of devices, velocity of UAVs, etc. \\ \hline
\cite{doubleQ_application} &Data capture & Online double Q-learning& Battery level, queue length, etc. &  Select IoT nodes\\ \hline
\cite{RLCoverage} &  Trajectory design, coverage problem & DDPG & \makecell{Remaining energy, \\[-2pt] instantaneous position, etc.} & Flying distance and direction\\ \hline
\cite{DRLCoverage}  & Coverage, fairness, energy efficiency & DDPG & \makecell{Current coverage score, \\[-2pt]energy consumption, etc.}& Flight distance and direction\\ \hline
\cite{DDPG-RA,DDPG_MC}  & Online flight resource allocation & DDPG& Battery level, queue length, etc.& Adjust heading and velocity, node selection \\ \hline
\end{tabular}
}
\label{table:ML}
\end{center}
\end{sidewaystable}

In summary, the main lessons learned from this section include:
\begin{itemize}
    \item Machine learning tools such as supervised learning, CNN, RNN, SNN are being used for channel modeling, resource management, and positioning problems. 
    \item ML tools make these problems model-free and easier to analyze the consumer behavior and requirements.
    \item Machine learning methods are limited due to their high computational requirements.
    \item Machine learning-based methods may be combined with traditional optimization methods to better serve users.
    \item Federated learning and distributed learning are used to protect the privacy of data.
    \item Reinforcement learning enables an agent to learn by interacting with the dynamic environment. However, it also suffers from computational complexity.
\end{itemize}
\section{The Intersection of game theory and machine learning in U-WCNs}\label{sec:intersection}
With the increased deployment of mobile Internet and IoT systems, there are increasing communication requirements for ultra-Reliable Low Latency Communication (uRLLC), massive Machine-Type Communication (mMTC), and enhanced Mobile Broadband (eMBB) systems. 
UAVs have the potential of playing a major role in such fields and are also called upon in the elastic and reliable operation of V2X and Wireless Sensor Networks (WSN). 
It is important to delineate the limitations and benefits of deploying UAVs where ML and game-theoretic approaches may find broader applications. 
To be able to support massive wireless traffic demands, future networks will be multilayered and very dense. Consequently, a large number of UAVs will be deployed to satisfy such increasing demands, necessitating adaptive and data-driven algorithms. 

Swarms of UAVs equipped with innovative wireless communication technologies will be deployed to relay data, replace damaged communication infrastructures, assist overloaded networks, provide network backhaul, and to serve as flying base stations.
Due to their large number and complexity, and with the dynamic nature of UAV-assisted networks, such systems must possess self-organizing capabilities. 
Self-organizing wireless networks will enhance network coverage, increase network capacity, improve quality of service, decrease operational costs by eliminating human involvement in performing tasks, and enhance network reliability. 
However, having a large number of UAVs induces interference to the network, which necessitates distributed techniques that suit the nature and features of these networks.
The large size, complexity, and dynamic nature as well as the need for self-organization, pose challenges for centralized algorithms. 
Centralized approaches hinder scalability and induce backhaul network congestion, which limits the downlink and uplink data rates.
Thus, advanced distributed algorithms are needed to address the interference challenge in UAVs-assisted networks. 
Players in UAV-assisted networks use distributed interference management approaches.
These distributed algorithms will have multi-objective schemes that include optimizing the transmit power, the 3D locations, the azimuth and elevation angles of the UAVs’ antennas, the trajectory of the UAV, and the hover or flight times. 
Two famous examples are realizing Ultra-Reliable and Low-Latency Communications and Massive Machine Type Communications in UAV ecosystem.

\subsection{uRLLC and mMTC}
Realizing Ultra-Reliable and Low-Latency Communications (uRLLC) in UAV ecosystems faces a serious challenge, namely, the restricted frequency spectrum accessible for concurrent Air-to-Air (A2A), Ground-to-Air (G2A), Air-to-Ground (A2G), and Ground-to-Ground (G2G) communications.  
This is due to the fact that the implementation of A2G, G2A, and A2A links typically depends on devoted wireless communication channels. 
UAV-based communication systems have a rigorous demand for system resources in both the G2A and A2G links due to the essential provision of high data rate backhauling and the exchange of time-critical UAV controlling signals.
Moreover, in order to achieve URLLC, multiple UAVs should be deployed simultaneously, which places a substantial burden on the need for an available frequency spectrum. 
Henceforth, classical multiple access schemes based on orthogonal spectrum partition would rapidly drain the available resources even with small numbers of UAVs and ground users. 
Furthermore, this will lead to a long delay in achieving URLLC and introduces severe safety issues in controlling UAVs~\cite{GT_survey}. 
In this case, conventional model-based methods with ideal assumptions can not address these challenges.
Moreover, conventional optimization problems of resource management and schedule design are neither convex nor deterministic.
Deep learning may be an option for these non-convex and non-deterministic problems~\cite{URLLC_review}, but conventional data-driven deep learning is data-dependent and has a long training phase, which limits its applicability in real systems.
To make the best use of deep learning in URLLC,  well-established models may be combined with deep learning in order to reduce the latency of URLLC.
Transfer learning (model transfer) and federated learning can also be applied to reduce training cost and improve the learning efficiency.
On the other hand, a multi-level architecture which enables device intelligence, UAV intelligence, and BS intelligence may also be proposed.

Massive Machine Type Communications (mMTC) is another provisioned service of 5G and Beyond 5G (B5G). 
This service provides connections to a large number of devices/machines that sporadically exchange small amounts of data. 
In many practical data collection applications in mMTC networks, e.g., distributed intelligence realized by pervasive sensors, a large number of devices may be distributed in a wide area while each has to only transmit small bursts of data. 
In such a setup, it is very costly in terms of energy consumption to have the UAVs get close to each of them in order to collect data. 
This in turn will lead to an energy consumption adjustment between UAVs and ground devices \cite{Polo}. 
To tackle this problem, inter-UAV cooperation among UAV swarms is used, where the UAVs organize into clusters and dynamically select a cluster head based on some criteria relating the remaining energy and location-related physical parameters. 
The UAVs then collect data in a small area and transmit it to the UAV cluster head for further processing. 
Furthermore, in the case of large-scale deployment of UAVs in heterogeneous applications, the data exchange activities of UAVs to the same ground BS may be highly random, which requires more efficient random access protocols \cite{Hoefer}.
The performance of UAV-based mMTC faces yet another challenge, namely the small batteries in  UAVs due to their size, weight, and power limitations. 

In the following subsections, some potential solutions for solving the above problems based on a combination of game theory and machine learning methods and their respective challenges in U-WCNs are summarized.
Other challenges and open problems such as the softwarization~\cite{softwarization} of U-WCNs, intelligent reflective surface for UAV communications~\citep{Agyapong}, and effective routing protocols~\cite{U2RV}, are also important but are out of the scope of this article.

\subsection{Combining game theory and machine learning in U-WCNs}
Based on the review of existing works of game theory in U-WCNs (Section \ref{sec:GT}) and machine learning used in U-WCNs (Section \ref{sec:ML}), we classify three examples for combining game theory and machine learning methods for solving problems in U-WCNs.
\begin{itemize}
    \item One approach for combining game theory and machine learning in U-WCNs is to use a machine learning-based method to analyze the user's communication behavior and habits by collecting historical data, then use game theory to optimize a specific objective (such as association, positioning, trajectory planning, etc.), as done in Reference \cite{security,CacheESN,SNN-LSM}.
    \item Another approach in applications such as search and rescue and parcel delivery tasks, machine learning (CNN, RNN) may be used by UAVs for obstacle (victims, items, etc) recognition, then game theory is used to make some high-level decisions.
    \item Yet another unification of game theory and machine learning is found in Multi-Agent Reinforcement Learning (MARL)~\cite{zhang2021multiagent}.
    Multi-agent reinforcement learning involves the participation of more than one player in optimizing an objective.
    To be more specific, multiple players make decisions in a common environment and aim to maximize their own long-term return by interacting with the environment and other players.
    Without the need for exact modeling, MARL allows deep neural networks to be combined with game theory, thus realizing high-level decision making, as done in \cite{QL,zhang2021multiagent}.
\end{itemize}
In the following two subsections, we will introduce the benefits of mean field game, Evolutionary game and MARL in solving problems in multi-UAVs communication networks and their challenges, which we think are three tools that have great potentials in U-WCNs.
\subsubsection{Mean field game and Evolutionary game}
Mean field game and Evolutionary game are suitable for large scale networks and thus are believed to be appropriate options for interference management of massive UAVs~\cite{survey-UAV-2}.
As a special form of differential game, MFG models each player's interaction with the collective behavior (mean field) of all the players instead of each of them.
Such mean field approximation can thus be used to model the distribution of states (such as aggregated interference from other setups), which significantly simplifies the original problem which analyzes the coupling and gaming between every two players, thus reduces the computational complexity.
One recent example is \cite{RobustMFG} for minimizing delay and energy consumption in a UAVs-caching system.
In this work, a distributed delay optimization algorithm based on mean field game theory is proposed to model the large-scale UAVs caching and dynamic flight strategy problem. Simulation results show that the proposed algorithm has a larger delay reduction and higher average energy efficiency compared to other two strategies.

Evolutionary game theory provides a solid basis for games among multi-agents in an uncertain environment based on the intuition that in the real world, players are not completely rational and knowledgeable.
Recent applications of Evolutionary game theory in U-WCNs are mainly on access selection~\citep{jointaccessselect,EGT-modeselec}.
What's more, the study of Evolutionary game (i.e., population dynamics) so far is limited to a single population.
However, in the predictable future, many problems related to the interactions among different populations in the massive U-WCNs will appear.
It's expected that multi-UAVs applications in wireless communication will benefit from mean field game and evolutionary game perspectives.
\subsubsection{Multi-agent reinforcement learning}
Future U-WCNs are highly dense, dynamic and non-deterministic communication networks.
On one hand, due to the complexity of such systems, generating an exact model of the network environment is impractical. Model-free learning algorithms may however be used.  
On the other hand, the introduction of multiple intelligent agents results in a non-stationary environment, which makes the optimization/learning hard~\cite{matignon2012independent}. 
Hence, game theory and its solution frameworks are the necessary guidelines for creating stable algorithms. 
Multi-Agent Reinforcement Learning (MARL), at the intersection of game theory and machine learning, is a promising toolkit to solve problems in the dynamic and stochastic U-WCNs environments.

Recent studies of MARL assume that each agent is an independent learner, which means that each agent tries to optimize its behavior by receiving feedback from the environment but without communicating with other agents.

For example, \cite{MADDPG_independent_1} proposed a multi-agent DDPG (MADDPG) framework for solving UAVs' trajectory control problem in a UAV-aided mobile edge computing network.
In this article, each UAV learns its offloading decision and flying control independently in order to maximize the geographical fairness among the covered user equipments and the fairness of user equipment-load of each UAV, and minimize the overall energy consumption.
Reference \cite{MADDPG_independent_2} presented a MADDPG approach to jointly designing the UAVs' trajectory and allocating UAVs' transmission power in aims to satisfy the user equipments' quality of service requirement.
\cite{MADDPG_independent_3} proposed a multi-agent deep Q-learning method for multi-UAV trajectory design in a cellular Internet of UAVs.
Similarly, the UAVs need to determine its movements at each cycle to optimize the reward function which is a sum of valid transmission probability for the UAV in the UAV-to-Device (U2D) and the cellular modes.
All the above works are similar in that the formulated optimization problem is either non-convex, highly-coupled or stochastic, and is therefore hard to solve using traditional optimization methods while greedy search algorithms have a high time and space complexities.
In such case, multi-agent reinforcement learning solves these problems without exact knowledge of the exact model of the system.

However, the feedback from the environment is dependent on the joint actions taken by all the agents, which makes the problem non-stationary and state-dependent.
Multi-agent communication and cooperation are necessary to deal with the uncertainty of the dynamic environment.
Considering this point, \cite{Cooperative_MARL_1} proposed a centralized offline training and decentralized online decision making MADDPG mechanism for vehicle association and resource allocation in a UAV-assisted vehicular network.
In this mechanism, for the centralized offline training phase, the observations and actions of all the UAV agents are needed to train the network.
Reference \cite{Cooperative_MARL_2} considered a cellular internet of UAVs executing sensing tasks through cooperative sensing and transmission to minimize the Age of Information (AoI).
By selecting from a discrete set of tasks and a continuous set of locations for sensing and transmission cooperatively, UAVs are able to minimize the age of information.
Similarly, the authors regarded the whole UAV-task-Base station system as a dynamic environment, where the state includes the location of all UAVs, the amount of sensing data, the AoI of task, and so on.
Finally, a compound-action actor-critic algorithm where a deep Q-network is used to learn the task selection decision of UAVs and a DDPG is used for the sensing location selection, is proposed for this optimization problem.
In the above two examples, scalability turns out to be a problem when combining action spaces of all agents without any effective mechanism.
It is stated that before each agent makes decisions, the agent needs to be able to decide when/who to communicate and distinguish between important and un-important information~\cite{Chen2020,NiuYaru}.
As a result, graph attention multi-agent reinforcement learning is proposed as a potential solution for the scalability problem of classical MARL and has more practical meaning~\cite{Chen2020,NiuYaru} by encoding the observation-action information into fixed-size features for each agent regardless of the number of neighbors.
To the authors' best knowledge, few applications of these algorithms are used in U-WCNs.

MARL and its variants enable agents to share information and learn from the environment to improve performance.
It is envisioned that MARL will play a more and more important part for the uRLLC and mMTC in the U-WCNs.
\subsection{Summary}
Conventional game theory methods consider the interaction of each player with other players under some coupling of their cost function.
This coupling relationship increases the computational complexity with the number of players increases. 
Machine learning-based methods are dependent on historical data.
Federated learning, which allows model parameters instead of data to be shared is also limited due to heterogeneous data distribution.
Mean field games and evolutionary games, are useful tools for dealing with large number of agents in interference management and resource allocation, but fail to model the interaction between the environment and players.
Multi-agent reinforcement learning, which allows each agent to learn without a model of the environment focuses on independent learning for each agent.  However, in practice, the action taken by one agent affects the reward of opponent agents and the evolution of the state.
With many successful and empirical applications of MARL in U-WCNs, the theoretical understanding of MARL algorithms remains in its infancy.

However, the combination of two or more of these methods can significantly alleviate the shortcomings of each method and solve problems in U-WCNs more efficiently.
One recent example in UAV coverage control is \cite{MFG_DRL_UAVcontrol}, which fused mean field game with multi-agent deep reinforcement learning where the MFG is used to construct the HJB/FPK equation and the distribution of state is obtained through the neural network feature embedding method.
In this way, the authors solved the difficulties of using MFG (i.e., complicated calculation process, limited sensing range, etc.) in real applications.
With the development of better theoretical understanding of these
algorithms and more efficient computational tools, the combination of game theory and machine learning has a more promising future for applications in U-WCNs.

\section{Conclusion} \label{sec:conclusion}
With the increased deployment of 5G, tele-medicine, IoT, AR/VR, smart cities and transportation, there is an increased desire for reliable wireless communications and privacy protection. 
UAVs-assisted wireless communication systems are a potentially excellent candidate for providing such services.
This article reviewed the state-of-the-art applications of game theory and machine learning-based algorithms in UAVs-assisted wireless communication systems. 
Several challenges and future research directions were also illustrated in this article. 
In addition, we discussed the combined use of game theory and machine learning.
In the near future, UAVs may deliver your parcel from an online store, based on an order from your phone or any IoT device; you will enjoy fast internet surfing and share your video when mountaineering; UAVs will monitor public safety including viral outbreaks or natural disasters. 
The technologies reviewed in this paper will help in making such scenarios possible.

\bibliography{main}

\end{document}